\documentclass[journal]{IEEEtran}
\usepackage[nolist]{acronym}

\usepackage[english]{babel} 
\usepackage{graphicx}       
\usepackage{verbatim}       
\usepackage{amssymb}        
\usepackage[cmex10]{amsmath}        
\usepackage{amsxtra}        
\usepackage{amsthm} 
\usepackage{array}
\usepackage{url}
\usepackage[colorlinks]{hyperref}
\usepackage[caption=false,font=footnotesize]{subfig}

\usepackage{tikz}
\usetikzlibrary{calc,positioning}
\usetikzlibrary{spy}
\usepackage{dsfont}
\usepackage{bbm}
\usepackage{siunitx}
\usepackage{pgfplots}
\usepackage{bm}
\usepackage{booktabs}
\usepackage{cleveref}
\usepackage{mathtools}

\newcommand{\FER}{\text{FER}}
\newcommand{\BER}{\text{BER}}

\newcommand{\SNR}{\mathsf{SNR}}

\newcommand{\HX}[1]{\mathbb{H}\left(#1\right)}

\newcommand{\I}[1]{\mathds{I}\left(#1\right)}

\newcommand{\abs}[1]{\left\lvert#1\right\rvert}

\newcommand{\cN}{\mathcal{N}}
\newcommand{\cU}{\mathcal{U}}
\newcommand{\cX}{\mathcal{X}}

\renewcommand{\d}[1]{\mathrm{d}#1}

\newcommand{\cC}{\mathcal{C}}

\newcommand{\mA}{\boldsymbol{A}}

\newcommand{\mH}{\boldsymbol{H}}

\newcommand{\va}{\boldsymbol{a}}

\newcommand{\rB}{\mathsf{B}}

\newcommand{\rL}{\mathsf{L}}

\newcommand{\rU}{\mathsf{U}}

\newcommand{\rX}{\mathsf{X}}
\newcommand{\rY}{\mathsf{Y}}
\newcommand{\rZ}{\mathsf{Z}}

\newcommand{\rvecB}{\bm{\mathsf{B}}}

\newcommand{\ch}{\mathrm{ch}}
\newcommand{\Jf}[1]{\mathsf{J}\left(#1\right)}
\newcommand{\Jfi}[1]{\mathsf{J}^{-1}\left(#1\right)}

\theoremstyle{plain}

\newtheorem{definition}{Definition}
\newtheorem{example}{Example}

\pdfoutput=1

\usepackage{cite}
\pgfplotsset{compat=newest}
\newtheorem{remark}{Remark}
\newcommand\undermat[2]{%
  \makebox[0pt][l]{$\smash{\underbrace{\phantom{%
    \begin{matrix}#2\end{matrix}}}_{\text{$#1$}}}$}#2}

\newcommand{\muinf}{\mathsf{I}}
\newcommand{\Iev}[3]{\muinf^{\mathsf{E,#3}}_{V_{#1}\rightarrow C_{#2}}}
\newcommand{\vecIev}[2]{\bm{\muinf}^{\mathsf{E,#2}}_{V\rightarrow C_{#1}}}
\newcommand{\Iec}[3]{\muinf^{\mathsf{E,#3}}_{C_{#1}\rightarrow V_{#2}}}
\newcommand{\vecIec}[2]{\bm{\muinf}^{\mathsf{E,#2}}_{C\rightarrow V_{#1}}}
\newcommand{\Ich}[1]{\muinf^{\mathsf{ch}}_{#1}}
\newcommand{\vecIch}{\bm{\muinf}^{\mathsf{ch}}}
\newcommand{\Iapp}[2]{\muinf^{\mathsf{APP},#2}_{#1}}
\newcommand{\fv}[2]{\mathsf{f}^{\mathsf{V}}_{{#1},{#2}}}
\newcommand{\fc}[2]{\mathsf{f}^{\mathsf{C}}_{{#1},{#2}}}
\newcommand{\neighborof}[1]{\mathcal{N}\left(#1\right)}
\newcommand{\converge}[1]{C_{#1}}
\newcommand{\sigmach}[1]{\sigma_{\mathsf{ch}#1}}
    
\begin{document}
\begin{acronym}
\acro{AWGN}{additive white Gaussian noise}
\acro{APP}{a posteriori probability}
\acro{BEC}{binary erasure channel}
\acro{BICM}{Bit-interleaved coded modulation}
\acro{BMD}{bit-metric decoding}
\acro{biAWGN}{binary input additive white Gaussian noise}
\acro{BP}{belief propagation}
\acro{EXIT}{extrinsic information transfer}
\acro{i.i.d.}{independent and identically distributed}
\acro{LDPC}{low-density parity-check}
\acro{LLR}{log-likelihood ratio}
\acro{MI}{mutual information}
\acro{P-EXIT}{protograph EXIT}
\end{acronym}

\title{Protograph-Based LDPC Code Design\\ for Shaped Bit-Metric Decoding}

\author{Fabian Steiner, Georg B\"ocherer and Gianluigi Liva\thanks{Fabian Steiner was with the Institute for Communications Engineering, Technische Universit\"at M\"unchen. He is now with the Institute for Circuit Theory and Signal Processing, Technische Universit\"at M\"unchen. Georg B\"ocherer is with the Institute for Communications Engineering, Technische Universit\"at M\"unchen. Gianluigi Liva is with the Institute of Communication and Navigation of the Deutsches Zentrum f\"ur Luft- und Raumfahrt (DLR) in Wessling, Germany. Emails: \texttt{fabian.steiner@tum.de}, \texttt{georg.boecherer@tum.de}, \texttt{Gianluigi.Liva@dlr.de}}\thanks{This work was supported by the German Ministry of Education and Research in the framework of an Alexander von Humboldt Professorship.}\thanks{Parts of this work will be presented at the 2015 IEEE International Symposium on Information Theory (ISIT).}}

\IEEEoverridecommandlockouts

\maketitle

\begin{abstract}
A protograph-based low-density parity-check (LDPC) code design technique for bandwidth-efficient coded modulation is presented. The approach 
jointly optimizes the LDPC code node degrees and the mapping of the coded bits to the bit-interleaved coded modulation (BICM) bit-channels. 
For BICM with uniform input and for BICM with probabilistic shaping, binary-input symmetric-output surrogate channels for the code
design are used. The constructed codes for uniform inputs perform as good as the multi-edge type codes of Zhang and Kschischang (2013). For 8-ASK and 64-ASK
with probabilistic shaping, codes of rates 2/3 and 5/6 with blocklength 64800 are designed, which operate within \SI{0.63}{\decibel} and \SI{0.69}{\decibel}
of $\frac{1}{2}\log_2(1+\text{SNR})$ for a target frame error rate of $10^{-3}$ at spectral efficiencies of 1.38 and 4.25 bits/channel use, respectively.

\end{abstract}
\section{Introduction}
\label{sec:intro}
\ac{BICM} combines high order modulation with binary error correcting codes \cite{zehavi_bicm,caire_bicm}. This makes BICM
attractive for practical application and BICM is widely used in standards, e.g., in DVB-T2/S2/C2. At a BICM receiver, \ac{BMD} is
used \cite[Sec. II]{martinez_mismatched}. Achievable rates for \ac{BMD} were investigated for uniformly distributed inputs in \cite{martinez_mismatched} and for non-uniformly distributed bits in \cite{fabregas_bicm_shaping}. These results were generalized to non-uniformly distributed input symbols in \cite{boecherer_bmd} and
\cite{bocherer2014achievable}. BICM with probabilistic shaping can achieve rates very close to the capacity $\frac{1}{2}\log_2(1+\SNR)$ of the \ac{AWGN} channel \cite{bocherer2014achievable}. The aim of this work is the design of \ac{LDPC} codes for BICM with probabilistic shaping to approach $\frac{1}{2}\log_2(1+\SNR)$ at high spectral efficiencies.

The key aspect of designing \ac{LDPC} codes for \ac{BICM} are the unequal error protection of the LDPC
coded bits and the BICM bit-channels that are different for different bit-levels. A first approach is to take an existing LDPC code and to 
optimize the mapping of the coded bits to the BICM bit-levels. This was done, e.g.,
in \cite{li2005,lei2008,hager2014optimized}. A more fundamental approach is to directly incorporate the different bit-channels in the code design.
This is done in \cite{zhang_MET_journal}, where the authors use multi-edge type (MET) codes \cite{richardson_MET} to parameterize the different
bit-channels. They then extend the extrinsic information transfer charts (EXIT) \cite{ten_brink_exit1} to multiple dimensions to design 
codes for quadrature amplitude modulation with 16 signal points (16-QAM). As 16-QAM can be constructed
as the Cartesian product of two four point amplitude-shift keying (4-ASK) constellations, two different bit-channels are apparent. For
constellations with more than two different bit-channels, the authors of \cite{zhang_MET_journal} observe long runtimes of their multidimensional EXIT approach. Therefore, they
suggest a high-order extension based on nesting, i.e., starting from $m=2$ bit-levels, they successively extend their codes from $m$ to $m + 1$ bit-levels by optimizing in each step only the additional
bit-level.

In this work, we follow \cite{zhang_MET_journal} and jointly optimize the code structure and the mapping of the coded bits to the BICM bit-levels,
but propose a protograph-based design~\cite{thorpe_protograph}. In \cite{divsalar_error_floor_cm}, protographs are designed for coded modulation by a variable degree matched
mapping (VDMM). Each bit-level is associated to a specific protograph variable node following the waterfilling approach (i.e., assigning the most protected coded bits to the bit-levels with highest bit-channel capacities). Recently, a 
protograph-based coded modulation scheme was introduced in \cite{marinoni2010proto} by performing a one-to-one mapping between the constellation 
symbols and the codeword symbols of a non-binary protograph LDPC code. This requires the constellation order to match the field order on which the 
LDPC code is constructed. To the best of our knowledge, none of the above-mentioned approaches leads to a joint optimization of the binary LDPC 
protograph ensemble and the bit-mapping.

We propose the following code design. We represent each bit-channel by a biAWGN surrogate. Our surrogates reflect both the BICM bit-channels and the 
input distribution. 
For the surrogate channels, we employ \ac{P-EXIT} analysis \cite{liva_pexit} to determine the ensemble iterative convergence threshold by 
accounting for the different bit-channels associated with the protograph variable nodes. The protograph ensemble is optimized with respect to the 
threshold by differential evolution. We optimize codes both for uniformly distributed inputs and for the probabilistic shaping scheme proposed 
in \cite{boecherer_bmd}. Our optimized codes perform as good as the codes by Zhang and Kschischang \cite{zhang_MET_journal}. Moreover, our design approach can be applied to arbitrary large constellations with shaped inputs. Our codes of rates 2/3 and 5/6 with blocklength 64800 operate within \SI{0.63}{\decibel} and \SI{0.69}{\decibel}
of $\frac{1}{2}\log_2(1+\SNR)$ for a target frame error rate of $10^{-3}$ at spectral efficiencies of 1.38 and 4.25 bits/channel use, 
respectively.

This paper is organized as follows. In Sec.~\ref{sec:bmd}, we review LDPC coded modulation with probabilistic shaping. Sec.~\ref{sec:protograph} introduces the concept of protographs 
and explains the necessary adaptions for \ac{P-EXIT} analysis. We present our code design approach in 
Sec.~\ref{sec:code_design_via_surrogates}. In Sec.~\ref{sec:sim_results}, we discuss the performance of our codes for uniform and shaped inputs, respectively.
We conclude in Sec.~\ref{sec:conclusion}.

\section{Preliminaries}
\label{sec:bmd}
\subsection{Channel Model}
Consider the discrete time \ac{AWGN} channel
\begin{align}
\rY = \Delta \rX+\rZ\label{eq:awgnchannel}
\end{align}
where the noise term $\rZ$ is zero mean, unit variance Gaussian, and where the input $\rX$ is distributed on the normalized $2^m$-ASK constellation
\begin{align}
\cX=\{\pm 1,\pm 2,\dotsc,\pm (2^m-1)\}.
\end{align}
The constellation spacing $\Delta$ controls the average power $\mathsf{E}[|\Delta 
\rX|^2]$, where $\mathsf{E}[\cdot]$ denotes expectation. The 
signal-to-noise ratio ($\SNR$) is $\SNR=\mathsf{E}[\abs{\Delta 
\rX}^2]/1$. Each signal point $x\in \cX$ is labeled by $m$ bits $\rvecB = (\rB_1, \rB_2,\dotsc, \rB_m)$, where $\rB_i$ denotes the $i$-th bit-level. Throughout this work, we label by the binary reflected Gray code (BRGC) \cite{gray1953pulse}, e.g., see Fig.~\ref{fig:8ASK_constellation_labeling}. Let $p_{\rY|\rvecB}$ be the transition function of the memoryless channel with input $\rvecB$ and output $\rY$. At the receiver, a soft-demapper calculates for each bit-level $i$ the soft-information
\begin{align}
L_i=\underbrace{\log\frac{p_{\rY|\rB_i}(y|0)}{p_{\rY|\rB_i}(y|1)}}_{\text{channel likelihood}}+\underbrace{\log\frac{P_{\rB_i}(0)}{P_{\rB_i}(1)}}_{\text{input priors}}\label{eq:lvalue}
\end{align}
which can be calculated by means of the joint distribution
\begin{align}
P_{B_i}(b_i)p_{\rY|\rB_i}(y|b_i)=\sum_{\va\in\{0,1\}^m\colon a_i=b_i}p_{\rY|\rvecB}(y|\va)P_{\rvecB}(\va).\label{eq:pbiy}
\end{align}
A bit-metric decoder uses the soft-information $L_1, L_2,\dotsc, L_m$ to estimate the transmitted data. For the decoder, the channel appears as $m$ parallel bit-channels, see Fig.~\ref{fig:bitchannels}.
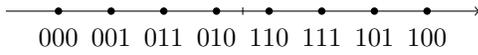
\begin{figure}
\centering
\begin{tikzpicture}[scale=.7]
 \draw[->] (-1,0) -- (8,0);
 \foreach \a/\b in {$000$/0, $001$/1, $011$/2, $010$/3, $110$/4, $111$/5, $101$/6, $100$/7} {
  \fill[black] (\b,0) circle (2pt);
  \node at (\b,-.5) {\a};
 }
 \draw (3.5,-2pt) -- (3.5,2pt);
\end{tikzpicture}
\caption{8-ASK constellation with BRGC \cite{gray1953pulse} labeling.}
\label{fig:8ASK_constellation_labeling}
\end{figure}
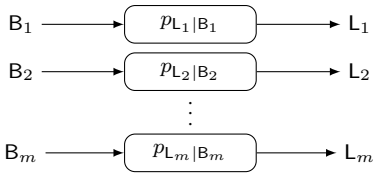
\begin{figure}
\centering
\footnotesize
\begin{tikzpicture}[scale=.9]
\node(B1)at (-5,5){$\rB_1$};

\path (B1) --++ (2.5,0) node (ch1)  [rectangle,draw,minimum width=1.75cm,minimum height=0.5cm,rounded corners] {$p_{\rL_1|\rB_1}$} --++(2.5,0) node 
(L1){$\rL_1$};

\path (B1) --++(0,-0.7) node (B2){$\rB_2$} --++ (2.5,0) node (ch2) [rectangle,draw,minimum width=1.75cm,minimum height=0.5cm,rounded corners,minimum height=0.5cm] {$p_{\rL_2|\rB_2}$} --++(2.5,0) node (L2){$\rL_2$};

\path (B2) --++(0,-1.2) node (Bm){$\rB_m$} --++ (2.5,0) node (chm) [rectangle,draw,minimum width=1.75cm,minimum height=0.5cm,rounded corners] {$p_{\rL_m|\rB_m}$} --++(2.5,0) node (Lm){$\rL_m$};
\path (ch2) --++(0,-0.5) node{$\vdots$};

\draw[-latex] (B1) -- (ch1);\draw[-latex] (ch1) -- (L1);
\draw[-latex] (B2) -- (ch2);\draw[-latex] (ch2) -- (L2);
\draw[-latex] (Bm) -- (chm);\draw[-latex] (chm) -- (Lm);

\end{tikzpicture}
\caption{$m$ parallel bit-channels that are different for different bit-levels.}
\label{fig:bitchannels}
\end{figure}
Bit-metric decoding can achieve the rate \cite[Theorem~1]{bocherer2014achievable}
\begin{align}
R_{\textnormal{BMD}}=\HX{\rvecB}-\sum_{i=1}^m \HX{\rB_i|\rY} = \HX{\rvecB}-\sum_{i=1}^m \HX{\rB_i|\rL_i}\label{eq:bmdrate}
\end{align}
where $\HX{\cdot}$ denotes entropy.
\begin{remark}
If the bits $\rB_1,\rB_2,\dots,\rB_m$ are independent, then \eqref{eq:bmdrate} can be written as \cite{martinez_mismatched,fabregas_bicm_shaping}
\begin{align}
R_{\textnormal{BMD}}=\sum_{i=1}^m\I{\rB_i;\rL_i}
\end{align}
where $\I{\cdot;\cdot}$ denotes mutual information.
\end{remark}

\subsection{LDPC Coded Modulation}\label{subsec:cm}

\subsubsection{LDPC codes}
\label{sec:ldpc_codes}
LDPC codes are linear block codes with a sparse $(n-k)\times n$ parity-check matrix $\mH$. The matrix $\mH$ can be represented by a Tanner 
graph~\cite[Section 3.3]{mct} consisting of variable nodes $V_i, i\in\{1,\ldots,n\}$ and check nodes $C_j, j\in\{1,\ldots,n-k\}$.  In the following, 
the set $\cN(V_i)$ denotes all neighbors of variable node $V_i$, i.e., all check nodes $C_j$ with a one occuring in the $i$-th column of $\mH$; the set
$\cN(C_j)$ is defined as all variable nodes $V_i$ possessing a one in the $j$-th row of $\mH$. The variable and check node degrees of the $i$-th variable and
$j$-th check node are then given as the cardinality of the sets $\cN(V_i)$ and $\cN(C_j)$, respectively.

\subsubsection{Transmitter}
\newcommand{\kd}{k_{\!\text{d}}}
\newcommand{\nc}{n_{\!\text{c}}}
\begin{figure*}
\footnotesize
\centering

\subfloat[][Canonical BICM model~{\cite[Sec. 2.3]{fabregas_ft}}.]{
\label{fig:canonical_bicm}
\begin{tikzpicture}
\draw (0,0) node (data) {$\rU^{\kd}$};
\draw node[draw,right=0.3cm of data, rectangle,rounded corners,minimum width=2cm,minimum height=0.5cm] (enc) {LDPC Encoder};
\draw node[draw,right=0.3cm of enc, rectangle,dashed,rounded corners,minimum width=2cm,minimum height=0.5cm] (pi) {Interleaver $\Pi$};
\draw node[draw,right=0.3cm of pi,rectangle,rounded corners,minimum width=2cm,minimum height=0.5cm] (bm) 
{Bit-Mapper};
\draw node[right=0.3cm of bm] (x) {$\rX^{\nc}$};
\draw[->] (data) -- (enc);
\draw[->] (enc) -- (pi);
\draw[->] (pi) -- (bm);
\draw[->] (bm) -- (x);
\end{tikzpicture}
}

\subfloat[][BICM with probabilistic shaping~\cite{boecherer_bmd}.]{
\label{fig:bicm_prob_shaping}
\begin{tikzpicture}
\draw (0,0) node (data) {$\rU^{\kd}$};
\draw node[draw,right=0.3cm of data, rectangle,rounded corners,minimum 
width=2cm,minimum height=.5cm] (matcher) {Matcher};
\draw node[right=0.3cm of matcher] (ms) {$(\rB_2\dotsb \rB_m)^{\nc}$};
\draw node[draw,right=0.3cm of ms, rectangle,rounded corners,minimum 
width=2cm,minimum 
height=0.5cm] (enc) {LDPC 
Encoder};
\draw node[right=0.3cm of enc] (encs) {$(\rB_2\dotsb 
\rB_m)^{\nc}\rB_{1}^{\nc}$};
\draw node[draw,right=0.3cm of encs,rectangle,rounded corners,minimum 
width=2cm,minimum 
height=0.5cm] (bm) 
{Bit-Mapper};
\draw node[right=0.3cm of bm] (x) {$\rX^{\nc}$};
\draw[->] (data) -- (matcher);
\draw[->] (matcher) -- (ms);
\draw[->] (ms) -- (enc);
\draw[->] (enc) -- (encs);
\draw[->] (encs) -- (bm);
\draw[->] (bm) -- (x);
\end{tikzpicture}}
\caption{Fig. (a) shows the canonical BICM model of \cite[Sec. 2.3]{fabregas_ft}, whereas Fig. (b) depicts the adapted model
needed for probabilistic shaping as proposed in \cite{boecherer_bmd}: Independent  
uniformly distributed data bits $\rU^{\kd}=\rU_1\rU_2\dotsb \rU_{\kd}$ are 
matched to $\nc$ strings $(\rB_2\rB_3\dotsb \rB_m)_i$, $i=1,2,\dotsc,\nc$ that 
are distributed according to $P_{\rB_2\dotsb \rB_m}$. The systematic LDPC encoder appends $\nc$ check bits 
$\rB_1^{\nc}=\rB_{11}\rB_{12}\dotsb \rB_{1\nc}$. The bit mapper maps this 
bit stream to signal points according to the BRGC labeling. The overall rate is 
$\kd/{\nc}$ [bits/channel use] and the rate of the LDPC code is 
$(m-1)\nc/(m\nc)=(m-1)/m$. At the receiver, a bit-metric decoder calculates an 
estimate of $(\rB_2\dotsb \rB_m)^{\nc}\rB_1^{\nc}$. A data estimate 
$\hat{\rU}^{\kd}=\hat{\rU}_1\hat{\rU}_2\dotsb \hat{\rU}_{\kd}$ is obtained 
by passing the estimate of $(\rB_2\dotsb \rB_m)^{\nc}$ through a dematcher. For the matcher and 
the dematcher, we use \cite{schulte2015constant}.}
\label{fig:ps}
\end{figure*}
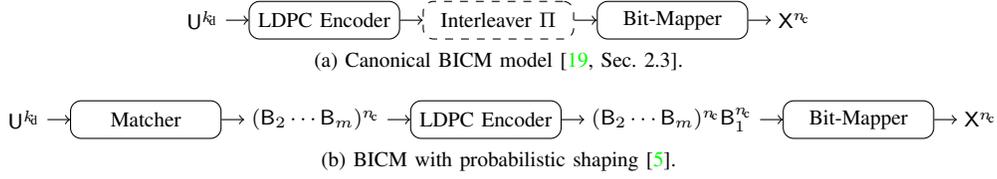
The canonical model of a BICM transmitter \cite[Sec.~2.3]{fabregas_ft} is displayed in Fig.~\ref{fig:canonical_bicm}. $k$ bits are encoded by a 
rate $k/n$ encoder. The coded bits are then permuted by an interleaver and a mapper modulates the bits to a sequence of channel input symbols of length
$\nc = n/m$. We adapt this general BICM model as follows. We describe the mapping function by specifying to which bit-level each coded bit is 
mapped. The mapper input is
\begin{align*}
(\rB_2\dotsb \rB_m)^{\nc}\rB_{1}^{\nc}
\end{align*}
where
\begin{align*}
&(\rB_2\dotsb \rB_m)^{\nc}=(\rB_2\dotsb \rB_m)_1(\rB_2\dotsb \rB_m)_2\dotsb (\rB_2\dotsb \rB_m)_{\nc}\\
\text{and}\\
&\rB_{1}^{\nc}=\rB_{11}\rB_{12}\dotsb\rB_{1\nc}.
\end{align*}
We put the bits for bit-level 1 in the end. At this point, this choice is arbitrary, but it will become useful later on when we consider shaped inputs. The memoryless modulation function of the mapper is
\begin{align}
B_{1i}B_{2_i}\dotsb B_{mi}\mapsto x_{B_{1i}B_{2i}\dotsb B_{mi}}.
\end{align}
For LDPC coded BICM, each coded bit has a variable node degree and it is transmitted over one of $m$ bit-channels. The resulting mapping of variable node degrees to bit-levels strongly influences the error performance of the system \cite{li2005}. If the LDPC code is given and fixed, we have to choose the interleaver appropriately. Since we are going to design new LDPC codes, we instead remove the interleaver and account for the mapping of variable node degrees to bit-levels in our code design. We choose a systematic rate $(m-1)/m$ LDPC encoder, which realizes the mapping
\begin{align}
(\rB_2\dotsb \rB_m)^{\nc}\mapsto (\rB_2\dotsb \rB_m)^{\nc}\rB_1^{\nc}.
\end{align}
Our model can be extended to code rates higher than $(m-1)/m$, see \cite{bocherer2015bandwidth}. For clarity of exposure, we only consider rate $(m-1)/m$ codes in the present work.

\subsubsection{Shaped Inputs} We use the shaping scheme proposed in \cite{bocherer2015bandwidth}. The capacity-achieving distribution of 
ASK constellations in Gaussian noise is symmetric around the origin. 
Consequently, it induces a distribution 
$P_{\rB_1\rB_2\dotsb \rB_m}$ on the BRGC labeling with the following properties:
\begin{itemize}
\item Bit-level $\rB_1$ decides on the sign of the transmitted constellation point, see 
Fig.~\ref{fig:8ASK_constellation_labeling}. Because of symmetry, bit-level $B_1$ is uniformly distributed.
\item Bit-levels $(\rB_2\dotsb \rB_m)$ and bit-level $\rB_1$ are independent. Bit-levels $B_2,\dotsc,B_m$ are correlated.
\end{itemize}
The scheme \cite{bocherer2015bandwidth} mimics the capacity-achieving distribution in the following way: 
first, generate bit-levels $\rB_2\dotsb \rB_m$ according to $P_{\rB_2\dotsb 
\rB_m}$, e.g., by using a distribution matcher (see 
Fig.~\ref{fig:bicm_prob_shaping}). In our implementation, we use the distribution matcher proposed in \cite{schulte2015constant}. The systematic encoder copies the bits $(\rB_2\dotsb \rB_m)^{\nc}$ to its output and leaves their distribution un-changed. The encoder appends check bits $\rB_1^{\nc}$ that are approximately uniformly distributed because each check bit is a modulo two 
sum of many information bits \cite[Sec.~7.1]{boecherer_thesis}. The signal point $x_{\rB_1\rB_2\dotsb \rB_m}$ 
selected by the bit-mapper then has approximately the capacity-achieving distribution.


\section{Protographs and EXIT Analysis for Parallel Channels}
\label{sec:protograph}

\subsection{Protograph-Based LDPC codes}

LDPC code ensembles as introduced in~\ref{sec:ldpc_codes} are usually characterized by the degree profiles of the variable and check nodes. 
For instance, $\lambda(x)=\sum_{d=1}^{d_v} \lambda_d x^{d-1}$ and $\rho(x)=\sum_{d=1}^{d_c} \rho_d x^{d-1}$ are the edge-perspective variable and 
check node degree polynomials with maximum degree $d_v$ and $d_c$, respectively. However, the degree profiles do not allow to characterize
the mapping of variable nodes to the $m$ different bit-channels resulting from our adapted BICM transmission scheme. In the following, we use 
protographs to incorporate the bit-mapping in our threshold analysis.

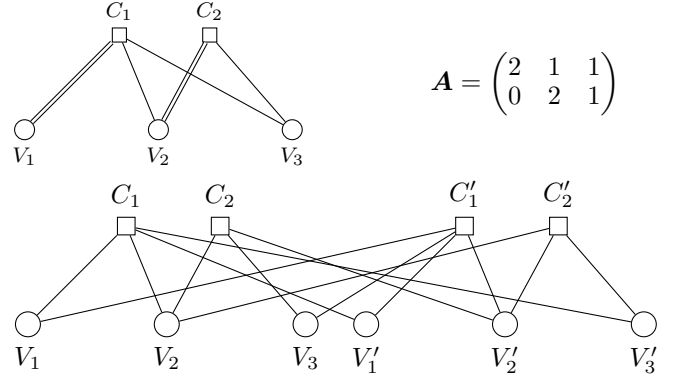
\begin{figure}
\begin{minipage}{.3\textwidth}
\footnotesize
\begin{tikzpicture}
 \node[rectangle,draw,thin,label={above:$C_1$}] (C1) {};
 \node[rectangle,draw,thin,label={above:$C_2$}] (C2) [right=of C1] {};
 
 \node[circle,draw,thin,label={below:$V_1$}] (V1) [below left=1.5cm of C1] {};
 \node[circle,draw,thin,label={below:$V_2$}] (V2) [right=1.5cm of V1] {};
 \node[circle,draw,thin,label={below:$V_3$}] (V3) [right=1.5cm of V2] {};

 \draw[double distance=1pt] (V1) -- (C1);
 \draw (V2) -- (C1);
 \draw[double distance=1pt] (V2) -- (C2);
 \draw (V3) -- (C1);
 \draw (V3) -- (C2);
\end{tikzpicture}
\end{minipage}
\begin{minipage}[b]{.1\textwidth}
 \begin{equation*}
 \boldsymbol{A} = \begin{pmatrix}
        2 & 1 & 1\\
        0 & 2 & 1
       \end{pmatrix}\label{eq:protograph_basematrix}
\end{equation*}
\end{minipage}

\begin{tikzpicture}
 \node[rectangle,draw,thin,label={above:$C_1$}] (C1) {};
 \node[rectangle,draw,thin,label={above:$C_2$}] (C2) [right=of C1] {};
 
 \node[circle,draw,thin,label={below:$V_1$}] (V1) [below left=1.5cm of C1] {};
 \node[circle,draw,thin,label={below:$V_2$}] (V2) [right=1.5cm of V1] {};
 \node[circle,draw,thin,label={below:$V_3$}] (V3) [right=1.5cm of V2] {};

\node[rectangle,draw,thin,label={above:$C_1'$}] (C1p) [right=3	cm of C2] {};
 \node[rectangle,draw,thin,label={above:$C_2'$}] (C2p) [right=of C1p] {};
 
 \node[circle,draw,thin,label={below:$V_1'$}] (V1p) [below left=1.5cm of C1p] {};
 \node[circle,draw,thin,label={below:$V_2'$}] (V2p) [right=1.5cm of V1p] {};
 \node[circle,draw,thin,label={below:$V_3'$}] (V3p) [right=1.5cm of V2p] {};

 \draw (V1) -- (C1);
 \draw (V2) -- (C1);
 \draw (V2) -- (C2);
 \draw (V3) -- (C2);

 \draw (V1p) -- (C1p);
 \draw (V2p) -- (C1p);
 \draw (V2p) -- (C2p);
 \draw (V3p) -- (C2p);

\draw (V1) -- (C1p);
\draw (V2) -- (C2p);
\draw (V1p) -- (C1);
\draw (V2p) -- (C2);

\draw (V3p) -- (C1);
\draw (V3) -- (C1p);

\end{tikzpicture}
\caption{Above, the Tanner graph of basematrix $\boldsymbol{A}$ with $M=2$ and $N=3$ is displayed. Below, an example lifting with $Q=2$ instances of the protograph is shown.}
\label{fig:protograph_base}
\end{figure}

Parity-check matrices are constructed from protographs as follows. Starting from a small bipartite graph represented via its basematrix $\boldsymbol{A}=[a_{lk}]$ of size $M\times N$, one applies a copy-and-permute operation (also known as lifting) to create $Q$ instances of the small graph and then permutes the edges so that the local edge connectivity remains the same. The $Q$ replicas of variable 
node $V_k, k\in\{1,\ldots,N\}$ must be connected only to replicas of the neighbors of $V_k$ while maintaining the original degrees for that specific edge. The resulting bipartite graph representing the
final parity-check matrix $\mH$ possesses $n = Q \cdot N$ variable nodes and $n-k = Q\cdot M$ check nodes. Parallel edges are allowed, but must be resolved during the copy-and-permute
procedure. An example protograph with the corresponding basematrix and an example lifting for $Q=2$ are shown in Figure~\ref{fig:protograph_base}.

\subsection{Protograph EXIT}

\begin{figure*}
\centering
\newcounter{myeq1}
\setcounter{myeq1}{\value{equation}}
\setcounter{equation}{5}
\setcounter{equation}{\value{myeq1}}
 \small
 \begin{align}
\Iev{k}{l}{\ell} &= \Jf{\sqrt{\sum_{\substack{l'=1\\l'\neq l}}^{M} a_{l'k} \cdot \Jfi{\Iec{l'}{k}{\ell-1}}^2 + (a_{lk}-1) \cdot \Jfi{\Iec{l}{k}{\ell-1}}^2 + \sigmach{,k}^2}}\label{eq:pexit_v2c}\\
\Iec{l}{k}{\ell} &= 1-\Jf{\sqrt{\sum_{\substack{k'=1\\k'\neq k}}^{N} a_{lk'} \Jfi{1-\Iev{k'}{l}{\ell}}^2 + (a_{lk}-1) \cdot \Jfi{1-\Iev{k}{l}{\ell}}^2}}\label{eq:pexit_c2v}\\
\Iapp{k}{\ell} &= \Jf{\sqrt{\sum_{l'=1}^{M} a_{l'k} \cdot \Jfi{\Iec{l'}{k}{\ell}}^2 + \sigmach{,k}^2}}.\label{eq:pexit_iapp}
 \end{align}
 \hrulefill
\end{figure*}

The performance of protograph-based \ac{LDPC} codes over parallel channels can be easily analyzed in the infinite blocklength limit by suitably 
modifying the \ac{P-EXIT}. We consider next the case where the codeword bits corresponding to the $N$ protograph variable nodes are transmitted
over $N$ binary-input symmetric-output parallel channels. In the following, we denote by $\Iev{k}{l}{\ell}$ the \ac{MI} between the message sent at 
iteration $\ell$ by the $k$-th variable node to the $l$-th check node and the corresponding codeword bit. Similarly, $\Iec{l}{k}{\ell}$ denotes the 
\ac{MI} between the message sent at iteration $\ell$ by the $l$-th check node to the $k$-th variable node and the corresponding codeword bit. We 
further express the \ac{MI} between the $k$-th channel output and input as $\Ich{k}$. The evolution of the \ac{MI} can be tracked by applying the 
recursion
\begin{align}
\Iev{k}{l}{\ell}&=\fv{k}{l}\left(\vecIec{k}{\ell-1},\Ich{k}\right) \label{eq:proto:Iev_general}\\
\Iec{l}{k}{\ell}&=\fc{k}{l}\left(\vecIev{l}{\ell}\right)\label{eq:proto:Iec_general}
\end{align}
with
\[
\vecIec{k}{\ell}:=\left(\Iec{1}{k}{\ell},\Iec{2}{k}{\ell},\ldots,\Iec{M}{k}{\ell} \right)
\]
 and
\[
\vecIev{l}{\ell}:=\left(\Iev{1}{l}{\ell},\Iev{2}{l}{\ell},\ldots,\Iev{N}{l}{\ell} \right)
\]
 where by convention we set $\Iev{k}{l}{\ell}=\Iec{l}{k}{\ell}=0$ if $C_l\notin \neighborof{V_k}$. In 
 \eqref{eq:proto:Iev_general}, \eqref{eq:proto:Iec_general} we introduced the variable and check \ac{EXIT} functions $\fv{k}{l}$ and 
 $\fc{k}{l}$, whose form depends on the underlying channel model.
 We moreover denote by $\Iapp{k}{\ell}$ the \ac{MI} between the \ac{APP} \ac{LLR} message computed at the $k$-th variable node in the 
 $\ell$-th iteration, and the corresponding codeword bit. Note that $\Iapp{k}{\ell}$ is a function of $\vecIec{k}{\ell-1}$ and $\Ich{k}$.

 Following \cite{Liva2013:Proto_TWCOM}, we define the protograph convergence region $\converge{\muinf}$ as the set of channel 
 \ac{MI} vectors $\vecIch=\left(\Ich{1},\Ich{2},\ldots,\Ich{N}\right)$ for which $\Iapp{k}{\ell}$ converges to $1$ for all 
 $k\in \left[1,N\right]$, as $\ell\rightarrow \infty$,
 \[
\converge{\vecIch}:=\left\{\vecIch\, \bigg| \, \Iapp{k}{\ell}\rightarrow 1, \, \forall k, \, \ell\rightarrow \infty   \right\}.
 \]
We will sometimes restate the convergence region with respect to other parameters characterizing the channel nuisance. 
Hence, we will denote the convergence region with respect to the generic channel parameter $\boldsymbol{\xi}$ as $\converge{\boldsymbol{\xi}}$. 

\begin{example}[Convergence region of parallel binary-input additive white Gaussian noise channels] 
\label{ex:biawgn_conv_region}
The convergence region of $N$ parallel 
\ac{biAWGN} channels,  with noise variances $\sigmach{,k}^2$ and inputs alphabets $\left\{\pm \sigmach{,k}^2/2  \right\}$, $k=1,\ldots,N$, may be 
expressed with respect to the noise standard deviation as
\[
\converge{\boldsymbol{\sigmach{}}}:=\left\{\sigmach{,1},\sigmach{,2},\ldots,\sigmach{,N} \bigg| \, \Iapp{k}{\ell}\rightarrow 1, \, \forall k, \, \ell\rightarrow \infty   \right\}.
\]
The evaluation of \eqref{eq:proto:Iev_general} and \eqref{eq:proto:Iec_general} at the $\ell$-th iteration is given by \eqref{eq:pexit_v2c} and 
\eqref{eq:pexit_c2v}, with $\Iapp{k}{\ell}$ evaluated as in  \eqref{eq:pexit_iapp} and where $\Jf{\sigma_\ch}$ is
\begin{equation*}
 \Jf{\sigma_{\ch}} := 1 - \int\limits_{-\infty}^\infty \frac{1}{\sqrt{2\pi\sigma^2_{\ch}}} \mathrm{e}^{-\frac{\left(z-\sigma^2_{\ch}/2\right)^2}{2\sigma^2_{\ch}}}\log_2\left(1+\mathrm{e}^{-z}\right)\d{z}
\end{equation*}
\end{example}

\begin{example}[Convergence region of parallel binary erasure channels]  
If the underlying parallel channels are  \acp{BEC} with erasure probabilities $\epsilon_k$, $k=1,\ldots,N$, the convergence region in terms of 
erasure probability is denoted by
\[
\converge{\boldsymbol{\epsilon}}:=\left\{\epsilon_1,\epsilon_2,\ldots,\epsilon_N \bigg| \, \Iapp{k}{\ell}\rightarrow 1, \, \forall k, \, \ell\rightarrow \infty   \right\}.
 \]
 Here, the recursion \eqref{eq:proto:Iev_general}, \eqref{eq:proto:Iec_general} at the $\ell$-th iteration is given by 
\begin{align*}
\Iev{k}{l}{\ell} &= 1- \epsilon_k \left(1-\Iec{l}{k}{\ell-1}\right)^{a_{lk}-1}\prod_{\substack{l'=1 \\l'\neq l}}^{M} \left(1-\Iec{l'}{k}{\ell-1}\right)^{a_{l'k}}\\
\Iec{l}{k}{\ell} &= \left(\Iev{k}{l}{\ell}\right)^{a_{lk}-1}\prod_{\substack{k'=1 \\k'\neq k}}^{N} \left(\Iev{k'}{l}{\ell}\right)^{a_{lk'}}  
\end{align*}
with
\[
\Iapp{k}{\ell} = 1- \epsilon_k \prod_{\substack{l'=1}}^{M} \left(1-\Iec{l'}{k}{\ell-1}\right)^{a_{l'k}}.
\]
\end{example}

\begin{figure}
\centering
\footnotesize
 \input{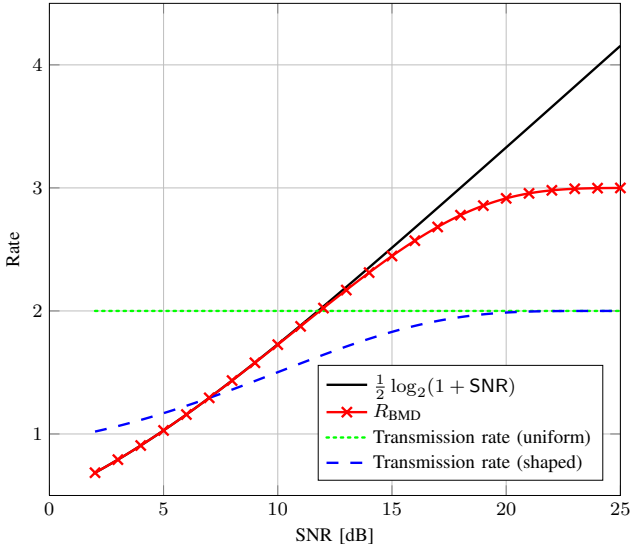}
 \caption{Search trajectory for protograph optimization for uniform and shaped inputs.}
 \label{fig:mi_rate}
\end{figure}

\section{Code Design via Surrogate Channels}
\label{sec:code_design_via_surrogates}

The $m$ bit-channels in Fig.~\ref{fig:bitchannels} are not output-symmetric and the input is not uniformly distributed when we perform shaping. Optimizing LDPC codes directly for the original bit-channels in Fig.~\ref{fig:bitchannels} is therefore difficult. To overcome this problem, we use the paradigm of \emph{universal codes}. A code is universal, if it performs similar on different channels, under certain conditions. Universality allows to replace the original channel by a surrogate channel for which code design is easier. We first derive a universality criterion and then construct surrogate channels.

\subsection{A Universality Criterion}

We start by considering the \emph{rate backoff} criterion \cite[Sec.~VIII.C]{bocherer2015bandwidth}. Let $R^*$ be the rate that can asymptotically be achieved by the considered code ensemble on the original channel and let $R$ be the actual transmission rate. The rate backoff is
\begin{align}
R^*-R.
\end{align}
In \cite[Sec.~VIII.C]{bocherer2015bandwidth}, it is assumed that the error performance is a function of $R^*-R$. Consider our transmission scheme of Fig.~\ref{fig:ps}. Suppose the code rate is $c$ so that $(1-c)m$ bits per channel use consist of uniformly distributed redundancy bits on average and suppose further that the input distribution is $P_{\rvecB}$. Then, by \eqref{eq:bmdrate} and Sec.~\ref{subsec:cm}, the achievable rate and the transmission rate, respectively, are given by 
\begin{align}
&R^*=\HX{\rvecB}-\sum_{i=1}^m\HX{\rB_i|\rL_i}\label{eq:achievablerate}\\
&R=\HX{\rvecB}-(1-c)m.\label{eq:transmissionrate}
\end{align}
Taking the difference, we have
\begin{align}
R^*-R=(1-c)m-\sum_{i=1}^m\HX{\rB_i|\rL_i}.
\end{align}
Note that the term $(1-c)m$ does not depend on the statistics, so that the sum of the conditional entropies determines the rate backoff. Since we want to account for the different bit-channels in our code design, we strengthen the rate backoff criterion and use as criterion the bit uncertainties
\begin{align}
\mathcal{U}=\{\HX{\rB_i|\rL_i}\}_{i=1}^m.\label{eq:uncertainties}
\end{align}
\subsection{Surrogate Channels}
\label{sec:surrogate_channels}

Suppose we optimized a code on some surrogate channel. If the performance of the code on the original channel cannot be improved by changing 
the bit-mapping, we succeeded to jointly optimize the code and the bit-mapping. In this case, we say that the surrogate channel is \emph{proper}. If in 
contrary the performance of the optimized code on the original channel can be improved by changing the bit-mapping, our joint optimization attempt 
failed. We will use the following two surrogate channels for code optimization. 

\subsubsection{BEC Surrogate} 
\label{sec:bec_surrogate}
We replace the original transmission scheme by a scheme where all input bits $\tilde{\rB}_i$ are uniformly 
distributed and where each bit-channel $P_{\tilde{\rL}_i|\tilde{\rB}_i}$ is a binary erasure channel (BEC) with erasure 
probability $\epsilon_i$. Criterion \eqref{eq:uncertainties} requires
\begin{align}
\HX{\tilde{\rB}_i|\tilde{\rL}_i}=\epsilon_i\overset{!}{=}\HX{\rB_i|\rL_i},\quad i=1,2,\dotsc,m.\label{eq:sur_match_bec}
\end{align}

\subsubsection{biAWGN Surrogate}
\label{sec:biawgn_surrogate}
All input bits $\tilde{\rB}_i$ are uniformly distributed and transmitted over biAWGN channels
\begin{align}
\tilde{\rL}_i=x_{\tilde{\rB}_i}+\rZ_i
\end{align}
where $x_{\tilde{\rB}_i}\in\{\pm\sigmach{,i}^2/2\}$ and the noise $\rZ_i$ has variance $\sigmach{,i}^2$. Applying 
criterion~\eqref{eq:uncertainties} yields
\begin{align}
\HX{\tilde{\rB}_i|\tilde{\rL}_i}\overset{!}{=}\HX{\rB_i|\rL_i},\quad i=1,2,\dotsc,m.\label{eq:sur_match_biawgn}
\end{align}
This can be achieved by numerically solving for $\sigmach{,i}$.

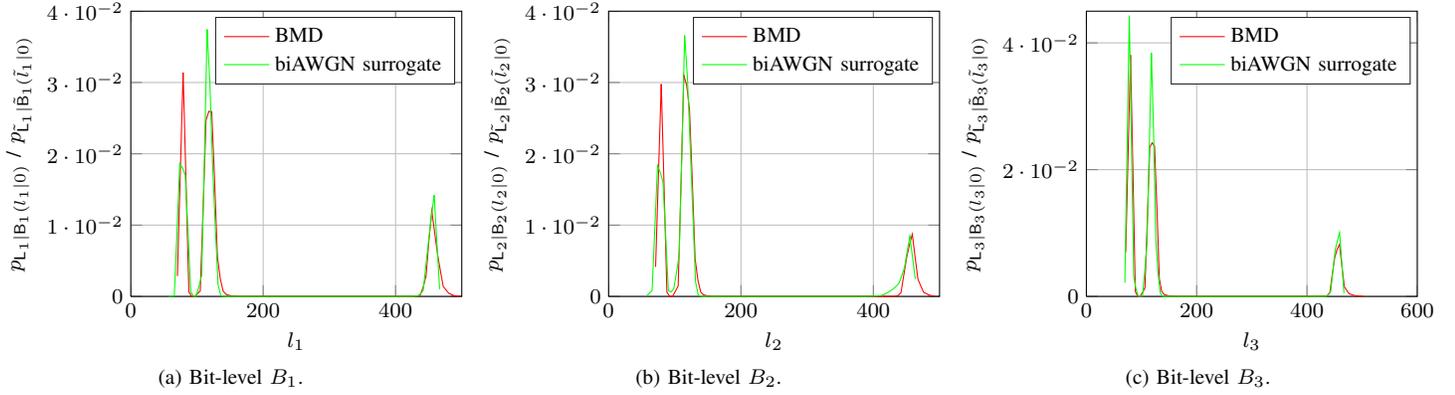
\begin{figure*}%
\centering
\footnotesize
\subfloat[][Bit-level $B_1$.\label{fig:llr_pdf_b1}]{
%
%
\definecolor{mycolor1}{named}{red}%
\definecolor{mycolor2}{named}{green}%
\begin{tikzpicture}

\begin{axis}[%
width=4.4cm,
at={(0.693333in,0.44in)},
scale only axis,
xlabel=$l_1$,
ylabel=$p_{\rL_1|\rB_1(l_1|0)}$ / $p_{\tilde{\rL}_1|\tilde{\rB}_1(\tilde{l}_1|0)}$,
xmin=0,
xmax=500,
xmajorgrids,
xminorgrids,
ymin=0,
ymax=0.04,
ymajorgrids,
yminorgrids,
legend style={legend cell align=left,align=left,draw=white!15!black}
]
\addplot [color=mycolor1,solid]
  table[row sep=crcr]{%
70.4053310640647	0.00285100085560415\\
74.7232305173466	0.0171228199947299\\
79.0411299706285	0.0313946391338557\\
83.3590294239104	0.0163230397797568\\
87.6769288771922	0.000640119207049286\\
91.9948283304741	2.51207397309013e-06\\
96.312727783756	5.1310447109926e-06\\
100.630627237038	0.000374576953578941\\
104.94852669032	0.000767946358411892\\
109.266426143602	0.0122414861263389\\
113.584325596884	0.024699929926541\\
117.902225050165	0.0259467417222827\\
122.220124503447	0.0259195044266085\\
126.538023956729	0.0164017478676882\\
130.855923410011	0.00555973280691954\\
135.173822863293	0.00249516290489682\\
139.491722316575	0.000726833862665074\\
143.809621769857	0.000307958889747887\\
148.127521223139	0.00015239202791648\\
152.445420676421	8.74736226459422e-05\\
156.763320129702	4.29511201016002e-05\\
161.081219582984	2.09410762267385e-05\\
165.399119036266	4.7034576517432e-06\\
169.717018489548	6.52070265355311e-07\\
174.03491794283	1.28276117774814e-07\\
178.352817396112	0\\
182.670716849394	0\\
186.988616302676	0\\
191.306515755958	0\\
195.624415209239	0\\
199.942314662521	0\\
204.260214115803	0\\
208.578113569085	0\\
212.896013022367	0\\
217.213912475649	0\\
221.531811928931	0\\
225.849711382213	0\\
230.167610835495	0\\
234.485510288777	0\\
238.803409742058	0\\
243.12130919534	0\\
247.439208648622	0\\
251.757108101904	0\\
256.075007555186	0\\
260.392907008468	0\\
264.71080646175	0\\
269.028705915032	0\\
273.346605368314	0\\
277.664504821595	0\\
281.982404274877	0\\
286.300303728159	0\\
290.618203181441	0\\
294.936102634723	0\\
299.254002088005	0\\
303.571901541287	0\\
307.889800994569	0\\
312.207700447851	0\\
316.525599901132	0\\
320.843499354414	0\\
325.161398807696	0\\
329.479298260978	0\\
333.79719771426	0\\
338.115097167542	0\\
342.432996620824	0\\
346.750896074106	0\\
351.068795527388	0\\
355.386694980669	0\\
359.704594433951	0\\
364.022493887233	0\\
368.340393340515	0\\
372.658292793797	0\\
376.976192247079	0\\
381.294091700361	0\\
385.611991153643	0\\
389.929890606925	0\\
394.247790060206	0\\
398.565689513488	0\\
402.88358896677	0\\
407.201488420052	0\\
411.519387873334	0\\
415.837287326616	0\\
420.155186779898	0\\
424.47308623318	0\\
428.790985686462	4.87449247544261e-06\\
433.108885139743	3.4730758887531e-05\\
437.426784593025	0.000216262844891856\\
441.744684046307	0.00130784984844961\\
446.062583499589	0.00287611090565857\\
450.380482952871	0.00786053601403455\\
454.698382406153	0.0120241543137989\\
459.016281859435	0.00896467269778127\\
463.334181312717	0.00596428361335187\\
467.652080765999	0.00362868550929046\\
471.969980219281	0.00140529693925254\\
476.287879672562	0.000902454557575266\\
480.605779125844	0.00041353013467656\\
484.923678579126	0.000251677743074188\\
489.241578032408	9.24229428567546e-05\\
493.55947748569	5.78525291164417e-05\\
497.877376938972	2.32821153761289e-05\\
};
\addlegendentry{BMD};

\addplot [color=mycolor2,solid]
  table[row sep=crcr]{%
57.4532590719135	1.32702064663326e-05\\
61.5854068754317	4.72052967396647e-05\\
65.71755467895	8.11403870129968e-05\\
69.8497024824683	0.00926964134420202\\
73.9818502859866	0.0186488624236185\\
78.1139980895049	0.0180494132401035\\
82.2461458930231	0.0170253397900809\\
86.3782936965414	0.00908203770985687\\
90.5104415000597	0.000687481588532769\\
94.642589303578	2.73446678700188e-05\\
98.7747371070963	5.4711676154964e-05\\
102.906884910615	0.00141942060276921\\
107.039032714133	0.00293610020192087\\
111.171180517651	0.019166389305598\\
115.303328321169	0.0374497402006042\\
119.435476124688	0.0300537671158716\\
123.567623928206	0.0183779067011826\\
127.699771731724	0.0097169187914452\\
131.831919535243	0.00164419868755204\\
135.964067338761	9.62201670883746e-05\\
140.096215142279	1.63085028963342e-05\\
144.228362945797	0\\
148.360510749316	0\\
152.492658552834	0\\
156.624806356352	0\\
160.756954159871	0\\
164.889101963389	0\\
169.021249766907	0\\
173.153397570425	0\\
177.285545373944	0\\
181.417693177462	0\\
185.54984098098	0\\
189.681988784498	0\\
193.814136588017	0\\
197.946284391535	0\\
202.078432195053	0\\
206.210579998572	0\\
210.34272780209	0\\
214.474875605608	0\\
218.607023409126	0\\
222.739171212645	0\\
226.871319016163	0\\
231.003466819681	0\\
235.1356146232	0\\
239.267762426718	0\\
243.399910230236	0\\
247.532058033754	0\\
251.664205837273	0\\
255.796353640791	0\\
259.928501444309	0\\
264.060649247828	0\\
268.192797051346	0\\
272.324944854864	0\\
276.457092658382	0\\
280.589240461901	0\\
284.721388265419	0\\
288.853536068937	0\\
292.985683872455	0\\
297.117831675974	0\\
301.249979479492	0\\
305.38212728301	0\\
309.514275086529	0\\
313.646422890047	0\\
317.778570693565	0\\
321.910718497083	0\\
326.042866300602	0\\
330.17501410412	0\\
334.307161907638	0\\
338.439309711157	0\\
342.571457514675	0\\
346.703605318193	0\\
350.835753121711	0\\
354.96790092523	0\\
359.100048728748	0\\
363.232196532266	0\\
367.364344335785	0\\
371.496492139303	0\\
375.628639942821	0\\
379.760787746339	0\\
383.892935549858	0\\
388.025083353376	0\\
392.157231156894	0\\
396.289378960413	0\\
400.421526763931	0\\
404.553674567449	0\\
408.685822370967	0\\
412.817970174486	0\\
416.950117978004	3.35106223897311e-07\\
421.082265781522	3.07180705239184e-06\\
425.21441358504	7.98669833621873e-06\\
429.346561388559	3.20696656269711e-05\\
433.478709192077	7.70632612889184e-05\\
437.610856995595	0.000357301426126757\\
441.743004799114	0.000837396942896926\\
445.875152602632	0.00438197185596302\\
450.00730040615	0.00791421485998974\\
454.139448209668	0.01115665800159\\
458.271596013187	0.0141984507065281\\
462.403743816705	0.00760902245167903\\
466.535891620223	0.00101959419682989\\
};
\addlegendentry{biAWGN surrogate};

\end{axis}
\end{tikzpicture}%
}
\subfloat[][Bit-level $B_2$.\label{fig:llr_pdf_b2}]{
%
%
\definecolor{mycolor1}{named}{red}%
\definecolor{mycolor2}{named}{green}%
\begin{tikzpicture}

\begin{axis}[%
width=4.4cm,
xlabel=$l_2$,
ylabel=$p_{\rL_2|\rB_2(l_2|0)}$ / $p_{\tilde{\rL}_2|\tilde{\rB}_2(\tilde{l}_2|0)}$,
at={(0.693333in,0.44in)},
scale only axis,
xmin=0,
xmax=500,
xmajorgrids,
xminorgrids,
ymin=0,
ymax=0.04,
ymajorgrids,
yminorgrids,
legend style={legend cell align=left,align=left,draw=white!15!black}
]
\addplot [color=mycolor1,solid]
  table[row sep=crcr]{%
71.031918999895	0.00414427185195455\\
75.2937559499022	0.0169590452266522\\
79.5555928999093	0.0297738186013498\\
83.8174298499165	0.0154727532729092\\
88.0792667999237	0.000606774638153291\\
92.3411037499309	2.52241358510572e-06\\
96.602940699938	5.15216391851381e-06\\
100.864777649945	0.000718780535006828\\
105.126614599952	0.00147877838136177\\
109.38845154996	0.0156521129793225\\
113.650288499967	0.031017744177426\\
117.912125449974	0.0296173752906991\\
122.173962399981	0.0263117791216051\\
126.435799349988	0.0162721973742653\\
130.697636299995	0.00529298973228651\\
134.959473250003	0.00228192560053425\\
139.22131020001	0.000598885387153084\\
143.483147150017	0.000227285564530271\\
147.744984100024	0.000111576549860314\\
152.006821050031	6.24163191378287e-05\\
156.268658000038	2.82295648035238e-05\\
160.530494950046	1.28804097962845e-05\\
164.792331900053	2.36140846265207e-06\\
169.05416885006	0\\
173.316005800067	0\\
177.577842750074	0\\
181.839679700082	0\\
186.101516650089	0\\
190.363353600096	0\\
194.625190550103	0\\
198.88702750011	0\\
203.148864450117	0\\
207.410701400125	0\\
211.672538350132	0\\
215.934375300139	0\\
220.196212250146	0\\
224.458049200153	0\\
228.719886150161	0\\
232.981723100168	0\\
237.243560050175	0\\
241.505397000182	0\\
245.767233950189	0\\
250.029070900196	0\\
254.290907850204	0\\
258.552744800211	0\\
262.814581750218	0\\
267.076418700225	0\\
271.338255650232	0\\
275.60009260024	0\\
279.861929550247	0\\
284.123766500254	0\\
288.385603450261	0\\
292.647440400268	0\\
296.909277350275	0\\
301.171114300283	0\\
305.43295125029	0\\
309.694788200297	0\\
313.956625150304	0\\
318.218462100311	0\\
322.480299050318	0\\
326.742136000326	0\\
331.003972950333	0\\
335.26580990034	0\\
339.527646850347	0\\
343.789483800354	0\\
348.051320750361	0\\
352.313157700369	0\\
356.574994650376	0\\
360.836831600383	0\\
365.09866855039	0\\
369.360505500397	0\\
373.622342450405	0\\
377.884179400412	0\\
382.146016350419	0\\
386.407853300426	0\\
390.669690250433	0\\
394.93152720044	0\\
399.193364150448	0\\
403.455201100455	0\\
407.717038050462	0\\
411.978875000469	0\\
416.240711950476	0\\
420.502548900483	0\\
424.764385850491	4.29346993209477e-07\\
429.026222800498	3.05909732661762e-06\\
433.288059750505	2.25943855176505e-05\\
437.549896700512	0.000143562900854423\\
441.811733650519	0.000553374605872908\\
446.073570600527	0.00303322917027671\\
450.335407550534	0.00544787666008685\\
454.597244500541	0.0072887018934725\\
458.859081450548	0.00871563662540416\\
463.120918400555	0.00548630321597907\\
467.382755350562	0.00239833230406818\\
471.64459230057	0.00147791968737533\\
475.906429250577	0.000586810002969061\\
480.168266200584	0.000384319227296635\\
484.430103150591	0.000184350865209321\\
488.691940100598	0.00010545835520708\\
492.953777050605	2.65658452048368e-05\\
};
\addlegendentry{BMD};

\addplot [color=mycolor2,solid]
  table[row sep=crcr]{%
57.5304727042027	8.83888791945916e-05\\
61.6294245808211	0.000454778639795405\\
65.7283764574394	0.000821168400396218\\
69.8273283340577	0.00948807745263458\\
73.9262802106761	0.0183279139901154\\
78.0252320872944	0.0173983799310097\\
82.1241839639127	0.0160531279741768\\
86.2231358405311	0.00863214499179758\\
90.3220877171494	0.000814918681665265\\
94.4210395937677	0.00061621110665775\\
98.5199914703861	0.00109470519699463\\
102.618943347004	0.00337535032424346\\
106.717895223623	0.00586078534205048\\
110.816847100241	0.0203914371925245\\
114.915798976859	0.0366028169731381\\
119.014750853478	0.0308917458753295\\
123.113702730096	0.0215269329644505\\
127.212654606714	0.0118441098120753\\
131.311606483333	0.00209923588087158\\
135.410558359951	0.000184366702562457\\
139.509510236569	3.12485936546547e-05\\
143.608462113188	0\\
147.707413989806	0\\
151.806365866424	0\\
155.905317743043	0\\
160.004269619661	0\\
164.103221496279	0\\
168.202173372898	0\\
172.301125249516	0\\
176.400077126134	0\\
180.499029002753	0\\
184.597980879371	0\\
188.696932755989	0\\
192.795884632608	0\\
196.894836509226	0\\
200.993788385844	0\\
205.092740262463	0\\
209.191692139081	0\\
213.290644015699	0\\
217.389595892318	0\\
221.488547768936	0\\
225.587499645554	0\\
229.686451522173	0\\
233.785403398791	0\\
237.884355275409	0\\
241.983307152028	0\\
246.082259028646	0\\
250.181210905264	0\\
254.280162781883	0\\
258.379114658501	0\\
262.478066535119	0\\
266.577018411738	0\\
270.675970288356	0\\
274.774922164974	0\\
278.873874041593	0\\
282.972825918211	0\\
287.071777794829	0\\
291.170729671448	0\\
295.269681548066	0\\
299.368633424684	0\\
303.467585301303	0\\
307.566537177921	0\\
311.665489054539	0\\
315.764440931158	0\\
319.863392807776	0\\
323.962344684394	0\\
328.061296561013	0\\
332.160248437631	0\\
336.259200314249	0\\
340.358152190868	0\\
344.457104067486	0\\
348.556055944104	0\\
352.655007820723	0\\
356.753959697341	0\\
360.852911573959	0\\
364.951863450578	0\\
369.050815327196	0\\
373.149767203814	0\\
377.248719080433	0\\
381.347670957051	0\\
385.446622833669	0\\
389.545574710288	1.50662862263499e-06\\
393.644526586906	9.70938445698152e-06\\
397.743478463524	2.01441826952316e-05\\
401.842430340143	4.20181982534893e-05\\
405.941382216761	6.77982880185787e-05\\
410.040334093379	0.000117014823024659\\
414.139285969998	0.000182971676060019\\
418.238237846616	0.000368900808305202\\
422.337189723234	0.00056292109426454\\
426.436141599853	0.000828143532908415\\
430.535093476471	0.00110073171148517\\
434.634045353089	0.00145618446430685\\
438.732997229708	0.00185348801220173\\
442.831949106326	0.00289250375121897\\
446.930900982944	0.00398229845492503\\
451.029852859563	0.00626539882881814\\
455.128804736181	0.00844063575354263\\
459.2277566128	0.00543842711817176\\
463.326708489418	0.00243621848280093\\
};
\addlegendentry{biAWGN surrogate};

\end{axis}
\end{tikzpicture}%
}
\subfloat[][Bit-level $B_3$.\label{fig:llr_pdf_b3}]{
%
%
\definecolor{mycolor1}{named}{red}%
\definecolor{mycolor2}{named}{green}%
\begin{tikzpicture}

\begin{axis}[%
width=4.4cm,
xlabel=$l_3$,
ylabel=$p_{\rL_3|\rB_3(l_3|0)}$ / $p_{\tilde{\rL}_3|\tilde{\rB}_3(\tilde{l}_3|0)}$,
at={(0.693333in,0.44in)},
scale only axis,
xmin=0,
xmax=600,
xmajorgrids,
xminorgrids,
ymin=0,
ymax=0.045,
ymajorgrids,
yminorgrids,
legend style={legend cell align=left,align=left,draw=white!15!black}
]
\addplot [color=mycolor1,solid]
  table[row sep=crcr]{%
71.3354260048363	0.00695246205696053\\
75.6909245134708	0.0225164875763905\\
80.0464230221053	0.0380805130958205\\
84.4019215307398	0.0197808632298384\\
88.7574200393742	0.000775720126660383\\
93.1129185480087	4.16329945108431e-07\\
97.4684170566432	8.50376058093818e-07\\
101.823915565278	0.000642645131652586\\
106.179414073912	0.00132626775960457\\
110.534912582547	0.0119752348183435\\
114.890411091181	0.0235100102709301\\
119.245909599816	0.0242275062118628\\
123.60140810845	0.0237157658446531\\
127.956906617085	0.0138872699526317\\
132.312405125719	0.00275876166179938\\
136.667903634353	0.0007787053009476\\
141.023402142988	0.00032339092842593\\
145.378900651622	0.000176054418277236\\
149.734399160257	8.88111495671726e-05\\
154.089897668891	5.01101808399658e-05\\
158.445396177526	2.23312296089013e-05\\
162.80089468616	1.00982156898641e-05\\
167.156393194795	1.85133954314176e-06\\
171.511891703429	0\\
175.867390212064	0\\
180.222888720698	0\\
184.578387229333	0\\
188.933885737967	0\\
193.289384246602	0\\
197.644882755236	0\\
202.000381263871	0\\
206.355879772505	0\\
210.71137828114	0\\
215.066876789774	0\\
219.422375298409	0\\
223.777873807043	0\\
228.133372315678	0\\
232.488870824312	0\\
236.844369332946	0\\
241.199867841581	0\\
245.555366350215	0\\
249.91086485885	0\\
254.266363367484	0\\
258.621861876119	0\\
262.977360384753	0\\
267.332858893388	0\\
271.688357402022	0\\
276.043855910657	0\\
280.399354419291	0\\
284.754852927926	0\\
289.11035143656	0\\
293.465849945195	0\\
297.821348453829	0\\
302.176846962464	0\\
306.532345471098	0\\
310.887843979733	0\\
315.243342488367	0\\
319.598840997002	0\\
323.954339505636	0\\
328.30983801427	0\\
332.665336522905	0\\
337.020835031539	0\\
341.376333540174	0\\
345.731832048808	0\\
350.087330557443	0\\
354.442829066077	0\\
358.798327574712	0\\
363.153826083346	0\\
367.509324591981	0\\
371.864823100615	0\\
376.22032160925	0\\
380.575820117884	0\\
384.931318626519	0\\
389.286817135153	0\\
393.642315643788	0\\
397.997814152422	0\\
402.353312661057	0\\
406.708811169691	0\\
411.064309678326	0\\
415.41980818696	0\\
419.775306695595	0\\
424.130805204229	2.39168266338878e-07\\
428.486303712864	1.54130660529505e-06\\
432.841802221498	2.44571697541337e-05\\
437.197300730132	0.000158143372553634\\
441.552799238767	0.000676332424870585\\
445.908297747401	0.00350153857429247\\
450.263796256036	0.00613689826874493\\
454.61929476467	0.007411691702583\\
458.974793273305	0.00820407388513163\\
463.330291781939	0.00475123705633288\\
467.685790290574	0.0015338126662831\\
472.041288799208	0.000964778212159265\\
476.396787307843	0.000422636900872628\\
480.752285816477	0.000292857113089998\\
485.107784325112	0.000166106790014326\\
489.463282833746	0.000110548887552196\\
493.818781342381	5.56464833015142e-05\\
498.174279851015	3.22079932003037e-05\\
502.52977835965	8.76950309909249e-06\\
};
\addlegendentry{BMD};

\addplot [color=mycolor2,solid]
  table[row sep=crcr]{%
69.785849026101	0.00215583441170811\\
73.7988167035681	0.0232198849768943\\
77.8117843810353	0.0442839355420805\\
81.8247520585024	0.0241747070999543\\
85.8377197359695	0.0032077020118426\\
89.8506874134366	0.00123534450567127\\
93.8636550909038	7.12698753271885e-05\\
97.8766227683709	4.02449343207354e-05\\
101.889590445838	8.31145382710841e-05\\
105.902558123305	0.00402621436930322\\
109.915525800772	0.00831600133163152\\
113.928493478239	0.0227875451652676\\
117.941461155707	0.0384161068251887\\
121.954428833174	0.0255938265505709\\
125.967396510641	0.00880166135485469\\
129.980364188108	0.00369937089410788\\
133.993331865575	0.000545392889189365\\
138.006299543042	1.78439078115827e-05\\
142.019267220509	2.76888224662489e-06\\
146.032234897976	0\\
150.045202575444	0\\
154.058170252911	0\\
158.071137930378	0\\
162.084105607845	0\\
166.097073285312	0\\
170.110040962779	0\\
174.123008640246	0\\
178.135976317713	0\\
182.148943995181	0\\
186.161911672648	0\\
190.174879350115	0\\
194.187847027582	0\\
198.200814705049	0\\
202.213782382516	0\\
206.226750059983	0\\
210.23971773745	0\\
214.252685414918	0\\
218.265653092385	0\\
222.278620769852	0\\
226.291588447319	0\\
230.304556124786	0\\
234.317523802253	0\\
238.33049147972	0\\
242.343459157187	0\\
246.356426834655	0\\
250.369394512122	0\\
254.382362189589	0\\
258.395329867056	0\\
262.408297544523	0\\
266.42126522199	0\\
270.434232899457	0\\
274.447200576924	0\\
278.460168254392	0\\
282.473135931859	0\\
286.486103609326	0\\
290.499071286793	0\\
294.51203896426	0\\
298.525006641727	0\\
302.537974319194	0\\
306.550941996661	0\\
310.563909674129	0\\
314.576877351596	0\\
318.589845029063	0\\
322.60281270653	0\\
326.615780383997	0\\
330.628748061464	0\\
334.641715738931	0\\
338.654683416398	0\\
342.667651093866	0\\
346.680618771333	0\\
350.6935864488	0\\
354.706554126267	0\\
358.719521803734	0\\
362.732489481201	0\\
366.745457158668	0\\
370.758424836135	0\\
374.771392513603	0\\
378.78436019107	0\\
382.797327868537	0\\
386.810295546004	0\\
390.823263223471	0\\
394.836230900938	0\\
398.849198578405	0\\
402.862166255872	0\\
406.87513393334	0\\
410.888101610807	0\\
414.901069288274	0\\
418.914036965741	0\\
422.927004643208	0\\
426.939972320675	3.36496106360638e-07\\
430.952939998142	3.63415794869514e-06\\
434.965907675609	2.91982478576344e-05\\
438.978875353077	0.000305259653515923\\
442.991843030544	0.000766605429510841\\
447.004810708011	0.00406897821733435\\
451.017778385478	0.00729605279072893\\
455.030746062945	0.00875361932504078\\
459.043713740412	0.0100838845752293\\
463.056681417879	0.00530368818749084\\
467.069649095346	0.00052349179975252\\
};
\addlegendentry{biAWGN surrogate};

\end{axis}
\end{tikzpicture}%
}
\caption{Densities of $L$-values for 8-ASK with shaped inputs after 100 iterations at $\SNR=\SI{8.1}{\decibel}$.}
\label{fig:llr_density_plot}
\end{figure*}

\subsection{Iterative Decoding Threshhold}
\label{sec:search_trajectory}

Since irregular LDPC codes have shown better performance than regular LDPC codes~\cite{richardson_design_irregular}, we want to allow different variable node degrees for each of the $m$ distinct bit-channels as well. In order to have up to $D$ different degrees per bit-channel, the protograph matrix $\mA$ must have at least $N=D\cdot m$ variable nodes. We introduce a mapping function of the form $T(k) = \left\lceil k/D\right\rceil$ to relate each variable node $V_k$ with a corresponding bit-level $T(k) \in\{1,\ldots,m\}$. We next define the iterative decoding threshholds for uniform and shaped inputs, respectively.
\subsubsection{Uniform Inputs}
\label{sec:search_trajectory_uniform}

For uniform inputs, the bit uncertainty set 
\begin{align}
\mathcal{U}=\{\HX{\rB_i|\rL_i}\}_{i=1}^m=\{\HX{\rB_i|\Delta x_{\rvecB}+Z}\}_{i=1}
\end{align}
is a function of the $\SNR$ via the constellation scaling $\Delta$. Let $\boldsymbol{\xi}^{\cU(\SNR)}$ denote the vector of surrogate channel 
parameters associated with the uncertainty set $\cU(\SNR)$. Given a protograph ensemble, its iterative decoding threshold is given by
\begin{align}
 \min \SNR \text{ s.t. } \boldsymbol{\xi}^{\cU(\SNR)} \in \converge{\boldsymbol{\xi}}\label{eq:trajectory}.
\end{align}
Thus, differential evolution~\cite{storn_DE} can then be employed to find the protograph ensemble with the lowest threshold.
We next instantiate the iterative decoding threshhold for BEC surrogates and biAWGN surrogates.
\begin{definition}[Iterative decoding threshold under \ac{BEC} surrogate]\label{def:dec_threshold_bec}
The iterative decoding threshold of a protograph under the \ac{BEC} surrogate is the minimum $\SNR$ on the original channel such that the resulting erasure probability 
vector $\boldsymbol{\varepsilon} = \left(\epsilon_{T(1)},\epsilon_{T(2)},\ldots,\epsilon_{T(N)}\right)$ of the BEC surrogates belongs to the protograph convergence 
region $\converge{\boldsymbol{\varepsilon}}$.
\end{definition}
\begin{definition}[Iterative decoding threshold under the \ac{biAWGN} surrogate]\label{def:dec_threshold_biawgn}
The iterative decoding threshold of a protograph under \ac{biAWGN} surrogate  is the minimum $\SNR$ on the original channel such that the resulting \ac{LLR} 
standard deviation vector $\boldsymbol{\sigmach{}} = \left(\sigmach{,{T(1)}},\sigmach{,{T(2)}},\ldots,\sigmach{,{T(N)}}\right)$ of the biAWGN surrogates belongs to the 
protograph convergence region $\converge{\boldsymbol{\sigmach{}}}$.
\end{definition}

\subsubsection{Shaped Input}
\label{sec:search_trajectory_shaped}
For shaped input, the definition of the iterative decoding threshold is ambiguous. Since both the input distribution $P_{\rvecB}$ and the constellation scaling 
$\Delta$ influence both $\SNR$ and the bit-uncertainty set, there are infinitely many bit-uncertainty sets corresponding to the same value of $\SNR$. We
use an information theoretic argument to uniquely parametrize the bit-uncertainty set as a function of $\SNR$. For each $\SNR$, we choose 
$P_{\rvecB}$ and $\Delta$ that maximize the achievable rate, i.e., we let $P_{\rvecB}(\SNR)$ and $\Delta(\SNR)$ be the solutions of
\begin{align}
\max_{\Delta,P_{\rvecB}\colon\mathsf{E}[|\Delta x_{\rvecB}|^2]\leq \SNR}\left\{\mathbb{H}(\rvecB)-\sum_{i=1}^m\mathbb{H}(\rvecB_i|x_{\rvecB}+\rZ)\right\}.
\end{align}
As in the uniform case, the bit uncertainty set is now parametrized by the $\SNR$ and we define the threshhold by \eqref{eq:trajectory}. Definitions~\ref{def:dec_threshold_bec} and \ref{def:dec_threshold_biawgn} now apply accordingly for shaped input. Note 
that for shaped input, we do not search along a constant rate but we search along the transmission rate. In Fig.~\ref{fig:mi_rate}, we display 
example search trajectories for uniform inputs and for shaped inputs for 8-ASK: In case of uniform inputs, the transmission rate is 2 bits/channel use for 
a rate $(m-1)/m$ code independently of the $\SNR$ (dotted curve), whereas the transmission rate is given by $\mathbb{H}(\rvecB)-1$ in case of
shaped input (dashed curve). In the following, we must keep in mind that for shaped input, changing the $\SNR$ also changes the transmission rate.

\section{Simulation Results of the Optimized Codes}
\label{sec:sim_results}

We choose the basematrix $\boldsymbol{A}$ from the set $\{0,1,\ldots,S\}^{M\times N}$. The parameters $M$ and $N$ are chosen
such that the desired code rate $c = (N-M)/N$ is met and the number of different degrees per bit-level $D$ (see \ref{sec:surrogate_channels}) is
guaranteed. The maximum number of parallel edges $S$ is crucial for both the performance of the code and the success of the optimization procedure. 
The product $M\cdot S$ describes the maximum variable node degree in the final code and thereby determines the size of the optimization search space. 
The optimization of the basematrices is performed as described in~\ref{sec:search_trajectory}.

\subsection{Properness of BEC and biAWGN Surrogates}

We designed codes using both BEC and biAWGN surrogates. Our simulations results reveal that BEC surrogates are not proper, i.e., 
the performance of the obtained codes on the original channel can be improved further by changing the bit-mapper. For the codes we obtained 
by using biAWGN surrogates, the resulting bit-mapper are optimal, i.e., changing the bit-mapping degrades the performance on the original 
channel. Thus, for the considered scenario, the biAWGN channel is a proper surrogate and can be used for code design.

This observation is also verified by comparing the densities of the $L$-values of both the original bit-channels and the biAWGN surrogates 
after decoding as illustrated in Fig.~\ref{fig:llr_density_plot}. The plot shows the densities $p_{\rL_i|\rB_i}(l_i|0)$ and 
$p_{\tilde{\rL}_i|\tilde{\rB}_i}(\tilde{l}_i|0)$ (see \ref{sec:biawgn_surrogate}) for 8-ASK with shaped input at $\SNR=\SI{8.1}{\decibel}$. We
observe a close match of the two densities.

To further illustrate the close matching of the biAWGN surrogates with the original channel, we also plot simulation results for the biAWGN 
surrogates, see the curves with triangle markers in Fig.~\ref{fig:bmd8}, Fig.~\ref{fig:bmd64} For each $\SNR$, we calculate the corresponding set of
surrogate channel parameters $\boldsymbol{\sigmach{}}^{\cU{(\SNR)}}$ and simulate transmission over the corresponding uniform input symmetric-output biAWGN 
channels.

\subsection{Code Design for Uniform Input}

\begin{table*}
 \centering
 \renewcommand{\arraystretch}{1.0}
 \footnotesize
 \caption{Optimized protographs and design parameters for uniform input.}
 \begin{tabular}{p{3cm}p{7cm}p{1.5cm}p{2.5cm}}
  Constellation (code rate) & Basematrix & \ac{P-EXIT} Decoding Threshold & Gap of Threshold to 
$R_\text{BMD}~\eqref{eq:bmdrate}$\\
  \toprule
  4-ASK (1/2) & $\boldsymbol{A} = \left(\begin{array}{llllll}
2 & 1 & 1 & 2 & 1 & 4\\
1 & 1 & 1 & 2 & 2 & 5\\
\undermat{B_2}{1 & 0 & 0} & \undermat{B_1}{1 & 0 & 6}
\end{array}\right)$\vspace{.5cm} & \SI{5.57}{\decibel} & \SI{0.28}{\decibel}\\
  4-ASK (3/4) & $\boldsymbol{A} = \left(\begin{array}{llllllll}
        1 & 1 & 1 & 1 & 6 & 6 & 1 & 1\\
        \undermat{B_2}{1 & 1 & 2 & 2} & \undermat{B_1}{6 & 6 & 2 & 2}
\end{array}\right)$\vspace{.5cm} &  \SI{9.57}{\decibel}& \SI{0.26}{\decibel}\\
 \end{tabular}
 \label{tab:results_uniform}
\end{table*}

\begin{table*}
 \centering
 \renewcommand{\arraystretch}{1.0}
 \footnotesize
 \caption{Optimized protographs and design parameters for shaped input.}
 \begin{tabular}{p{3cm}p{7cm}p{1.5cm}p{2.5cm}}
  Constellation (code rate) & Basematrix & \ac{P-EXIT} Decoding Threshold & Gap of Threshold to 
$\frac{1}{2}\log_2(1+\SNR)$\\
  \toprule
  8-ASK (2/3) & $\boldsymbol{A} = \left(\begin{array}{llllll}
        1 & 1 & 1 & 1 & 1 & 6\\
        \undermat{B_2}{2 & 2} & \undermat{B_3}{1 & 1} & \undermat{B_1}{2 & 6}
       \end{array}\right)$\vspace{.5cm} & \SI{7.74}{\decibel} & \SI{0.39}{\decibel}\\
\midrule
  64-ASK (5/6) & $\boldsymbol{A} = \left(\begin{array}{llllllllllll}
        2 & 2 & 2 & 1 & 2 & 2 & 6 & 2 & 2 & 0 & 6 & 6\\
        \undermat{B_2}{1 & 1} & \undermat{B_3}{1 & 2} & \undermat{B_4}{1 & 1} & \undermat{B_5}{6 & 1} & \undermat{B_6}{0 & 2} & \undermat{B_1}{6 & 6}
       \end{array}\right)$\vspace{.5cm} & \SI{25.52}{\decibel} & \SI{0.35}{\decibel}\\
 \end{tabular}
 \label{tab:results_shaped}
\end{table*}

\begin{figure}
 \centering
 \footnotesize
%
%
\begin{tikzpicture}[scale=.9]

\begin{axis}[%
scale only axis,
xmin=9.2,
xmax=10,
xlabel={$\SNR$ [dB]},
xmajorgrids,
xminorgrids,
ymode=log,
ymin=1e-06,
ymax=1,
yminorticks=true,
ylabel={BER/FER},
ymajorgrids,
yminorgrids,
legend style={at={(axis cs:9.34,2e-6)},anchor=south west,legend cell align=left,align=left,draw=white!15!black}
]
\addplot [color=blue,solid,line width=1.0pt,mark=+]
  table[row sep=crcr]{%
9.3	0.0660785032019972\\
9.4	0.0632292684250516\\
9.5	0.0496648757190926\\
9.6	0.0284872598502117\\
9.7	0.00754800528896914\\
9.8	0.00221117201538888\\
9.9	0.000155186803955793\\
10	8.48837301518254e-06\\
};
\addlegendentry{MET code \cite{zhang_MET_journal} BER};

\addplot [color=blue,dashed,line width=1.0pt,mark=+]
  table[row sep=crcr]{%
9.3	1\\
9.4	1\\
9.5	0.833333333333333\\
9.6	0.520833333333333\\
9.7	0.159090909090909\\
9.8	0.0462962962962963\\
9.9	0.00345065562456867\\
10	0.000363504180298073\\
};
\addlegendentry{MET code \cite{zhang_MET_journal} FER};

\addplot [color=red,solid,line width=1.0pt,mark=o,mark options={solid}]
  table[row sep=crcr]{%
9.3	0.0672118124041201\\
9.4	0.0599454854262547\\
9.5	0.0501212195923734\\
9.6	0.029268481992841\\
9.7	0.00915447622178392\\
9.8	0.0018607169381742\\
9.9	0.000150110492034352\\
10	7.55752937078782e-06\\
10.1	2.40036681352145e-06\\
};
\addlegendentry{Optimized Protograph BER};

\addplot [color=red,dashed,line width=1.0pt,mark=o,mark options={solid}]
  table[row sep=crcr]{%
9.3	1\\
9.4	1\\
9.5	0.916666666666667\\
9.6	0.638888888888889\\
9.7	0.2\\
9.8	0.0462962962962963\\
9.9	0.00339443312966735\\
10	0.000434593654932638\\
10.1	0.000179985601151908\\
};
\addlegendentry{Optimized Protograph FER};

\addplot [color=black,solid]
  table[row sep=crcr]{%
9.308	1e-06\\
9.308	0.01\\
};
\addlegendentry{4-ASK uniform BMD};

\end{axis}
\end{tikzpicture}%
 \caption{4-ASK, uniform inputs: Performance of a rate 3/4 optimized protograph code compared to the rate 3/4 MET code in \cite{zhang_MET_journal}.}
 \label{fig:zhang34}
\end{figure}
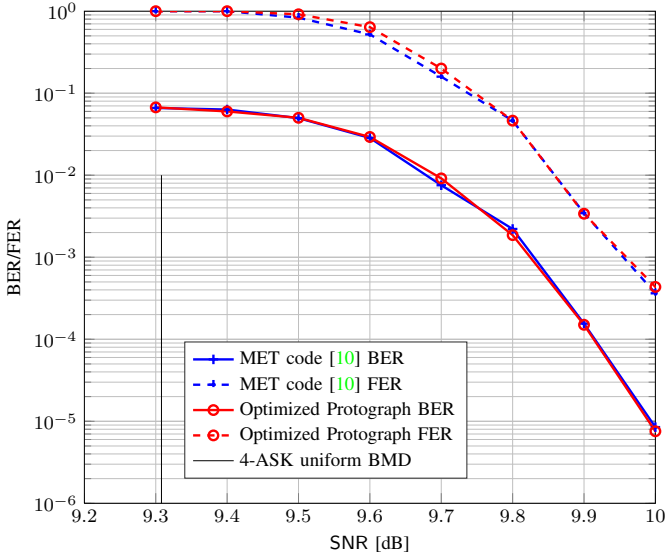

In order to compare our code design approach to the setting of Zhang and 
Kschischang~\cite{zhang_MET_journal}, we design codes of rates 1/2 and 
3/4 for 4-ASK constellations with uniform inputs. The 
optimized protographs are depicted in Table~\ref{tab:results_uniform}. We discuss the rate 1/2 code in more detail. The coded bits are transmitted over 
two different bit-channels. For each bit-channel, we allow $D=3$ (possibly different) variable degrees so that always 3 variable nodes in the 
protograph are assigned to the same surrogate channel: $\sigmach{,1} = \sigmach{,2} = \sigmach{,3}$ and $\sigmach{,4} = \sigmach{,5} = \sigmach{,6}$.
Once the optimized basematrices have been found, we construct quasi-cyclic parity-check matrices with blocklengths $n=16200$. We simulate the constructed codes using 100 decoding  
iterations. The bit error rates (BER) and frame error rates (FER) in Fig.~\ref{fig:zhang12} and \ref{fig:zhang34} show that the finite length performance of our codes is equal to or slightly 
better than the codes in \cite{zhang_MET_journal}. As a reference, we also plot the Shannon limit for BMD and a 4-ASK constellation for the corresponding code rates 1/2 and 3/4. The decoding thresholds given in Table~\ref{tab:results_uniform} 
are obtained by PEXIT analysis for the surrogate channels.

\begin{figure}
\centering
\footnotesize
%
%
\begin{tikzpicture}[scale=.9]

\begin{axis}[%
scale only axis,
xmin=5.1,
xmax=6.3,
xlabel={$\SNR$ [dB]},
xmajorgrids,
xminorgrids,
ymode=log,
ymin=1e-06,
ymax=1,
yminorticks=true,
ylabel={BER/FER},
ymajorgrids,
yminorgrids,
legend style={at={(axis cs:5.3,2e-6)},anchor=south west,legend cell align=left,align=left,draw=white!15!black}
]
\addplot [color=red,solid,line width=1.0pt,mark=+]
  table[row sep=crcr]{%
5.4	0.137416604892513\\
5.5	0.122621695082777\\
5.6	0.118477678939132\\
5.7	0.105618359278478\\
5.8	0.0478492298822173\\
5.9	0.0170130595136773\\
6	0.00459110658100651\\
6.1	0.000390074325710522\\
6.2	2.54981470022915e-05\\
6.3	0\\
6.4	0\\
};
\addlegendentry{MET code \cite{zhang_MET_journal} BER};

\addplot [color=red,dashed,line width=1.0pt,mark=+]
  table[row sep=crcr]{%
5.4	1\\
5.5	0.958333333333333\\
5.6	0.958333333333333\\
5.7	0.916666666666667\\
5.8	0.479166666666667\\
5.9	0.185185185185185\\
6	0.04375\\
6.1	0.00437445319335083\\
6.2	0.00025507601265177\\
6.3	0\\
6.4	0\\
};
\addlegendentry{MET code \cite{zhang_MET_journal} FER};

\addplot [color=blue,solid,line width=1.0pt,mark=o,mark options={solid}]
  table[row sep=crcr]{%
5.4	0.178343621399177\\
5.5	0.162289094650206\\
5.6	0.123533950617284\\
5.7	0.0579115226337448\\
5.8	0.0137288523090992\\
5.9	0.00218504489337823\\
6	0.000450824850191322\\
6.1	4.36436436436436e-05\\
6.2	3.44416891093157e-07\\
};
\addlegendentry{Optimized Protograph BER};

\addplot [color=blue,dashed,line width=1.0pt,mark=o,mark options={solid}]
  table[row sep=crcr]{%
5.4	1\\
5.5	1\\
5.6	1\\
5.7	0.555555555555556\\
5.8	0.194444444444444\\
5.9	0.0378787878787879\\
6	0.00730994152046784\\
6.1	0.000900900900900901\\
6.2	0.000169986401087913\\
};
\addlegendentry{Optimized Protograph FER};

\addplot [color=black,solid]
  table[row sep=crcr]{%
5.2805	1e-06\\
5.2805	0.01\\
};
\addlegendentry{4-ASK uniform BMD};

\end{axis}
\end{tikzpicture}%
 \caption{4-ASK, uniform inputs: Performance of a rate 1/2 optimized protograph code compared to the rate 1/2 MET code in \cite{zhang_MET_journal}.}
 \label{fig:zhang12}
\end{figure}
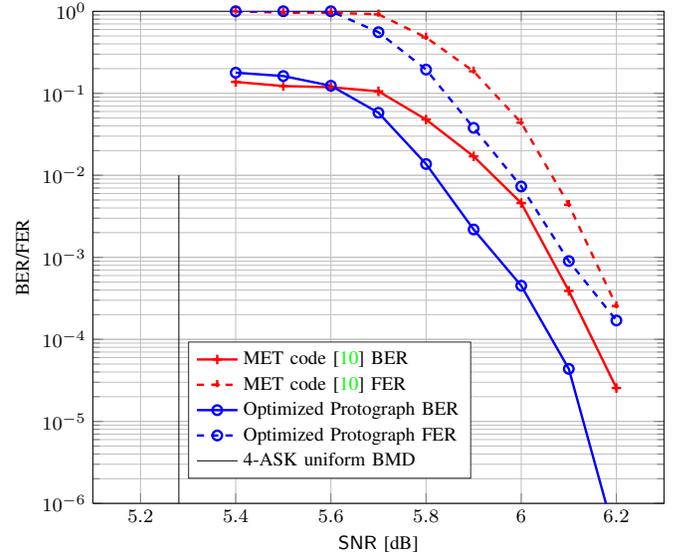

\subsection{Code Design for Shaped Input}

In the following we design codes of rate 2/3 for 8-ASK and rate 5/6 for 64-ASK to show that our design technique scales with the number of bit-levels. 
The optimized base matrices can be found in Table~\ref{tab:results_shaped}. In order to limit the search space for the optimization, we choose $D=2$ and allow a maximum of $S=6$ parallel edges. For comparison, we use the off-the-shelf DVB-S2 LDPC codes used in~\cite{bocherer2015bandwidth} and use the optimized bit-mappers
for DVB-S2 codes listed in \cite[Table V]{bocherer2015bandwidth}.

As the shaped input causes a non-constant transmission rate as shown in~\ref{sec:search_trajectory_shaped}, the bit and frame error ratios are plotted over the gap to continuous AWGN capacity, i.e.,
\begin{equation}
 \left.\Delta\SNR\right|_{\text{dB}} = \left.\SNR(\FER/\text{BER},\cC)\right|_{\text{dB}} - 10\log_{10}\left(2^{2\cdot R}-1\right)\label{eq:def_snr_gap},
\end{equation}
where $\left.\SNR(\FER/\BER,\cC)\right|_{\text{dB}}$ denotes the $\SNR$ in dB that is necessary to achieve the desired $\FER/\BER$ with code $\cC$ and the rate $R$ equals the transmission rate $\HX{\rvecB}-1$ at this specific $\SNR$. We use the same input distributions for our optimized codes and for the DVB-S2 codes.

\begin{figure}
 \centering
 \footnotesize
%
%
\begin{tikzpicture}[scale=.9]

\begin{axis}[%
scale only axis,
xmin=0.2,
xmax=0.9,
xlabel={$\Delta\SNR$ [dB]},
xmajorgrids,
xminorgrids,
ymode=log,
ymin=1e-07,
ymax=1,
yminorticks=true,
ylabel={BER/FER},
ymajorgrids,
legend style={at={(0.03,0.03)},anchor=south west,draw=black,fill=white,legend cell align=left}
]
\addplot [color=blue,solid,line width=1.0pt,mark=o]
  table[row sep=crcr]{%
0.420353421431698	0.0381602044753086\\
0.471599736806296	0.0305584490740741\\
0.522743412656327	0.0167303240740741\\
0.573791420494144	0.00781828703703704\\
0.624750881586509	0.00194404256687243\\
0.675629046610219	0.000360161968753414\\
0.72643579802999	3.007016221648e-05\\
0.777175050783995	1.52190323130494e-07\\
0.827854433158747	1.5e-07\\
0.878481681103471	1.34840501137102e-07\\
0.929064627961429	1.04493897556383e-07\\
0.979610715031387	9.43859251708385e-08\\
1.03012637912776	9.67455733001094e-08\\
1.08062164177263	7.48953656939768e-08\\
1.13110413065932	6.89548649036287e-08\\
1.18158162110742	8.36077736956498e-08\\
1.23206202937783	6.9816800714924e-08\\
};
\addlegendentry{DVB-S2 BER};

\addplot [color=blue,dashed,line width=1.0pt,mark=o,mark options={solid}]
  table[row sep=crcr]{%
0.420353421431698	1\\
0.471599736806296	0.979166666666667\\
0.522743412656327	0.7\\
0.573791420494144	0.398148148148148\\
0.624750881586509	0.142361111111111\\
0.675629046610219	0.0294985250737463\\
0.72643579802999	0.00794573643410853\\
0.777175050783995	0.00328731097961867\\
0.827854433158747	0.00274122807017544\\
0.878481681103471	0.00294204177699323\\
0.929064627961429	0.00237586124970302\\
0.979610715031387	0.00203873598369011\\
1.03012637912776	0.0021381227282446\\
1.08062164177263	0.00157828282828283\\
1.13110413065932	0.00148942508191838\\
1.18158162110742	0.00190803281816447\\
1.23206202937783	0.00157604412923562\\
};
\addlegendentry{DVB-S2 FER};

\addplot [color=red,solid,line width=1.0pt,mark=square]
  table[row sep=crcr]{%
0.265931622599248	0.079008487654321\\
0.317525865515046	0.080625\\
0.368997666751107	0.060883487654321\\
0.420353421431698	0.0374852109053498\\
0.471599736825913	0.0144023919753086\\
0.522743412656327	0.00597608024691358\\
0.573791420494144	0.000780396174863388\\
0.624750881666192	1.95841802168022e-05\\
0.675629046610219	2.92592333354072e-06\\
0.72643579802999	3.47425909630933e-07\\
};
\addlegendentry{Opt. BER};

\addplot [color=red,dashed,line width=1.0pt,mark=square,mark options={solid}]
  table[row sep=crcr]{%
0.265931622599248	1\\
0.317525865515046	1\\
0.368997666751107	0.958333333333333\\
0.420353421431698	0.694444444444444\\
0.471599736825913	0.433333333333333\\
0.522743412656327	0.175\\
0.573791420494144	0.0273224043715847\\
0.624750881666192	0.00127032520325203\\
0.675629046610219	9.99920006399488e-05\\
0.72643579802999	5.99952003839693e-05\\
};
\addlegendentry{Opt. FER};

\addplot [color=green,solid,line width=1.0pt,mark=triangle,mark options={solid}]
  table[row sep=crcr]{%
0.265931622599248	0.0798003472222222\\
0.317525865515046	0.0733931327160494\\
0.368997666751107	0.0579822530864198\\
0.420353421431698	0.0409458590534979\\
0.471599736825913	0.0177473958333333\\
0.522743412656327	0.00962384259259259\\
0.573791420494144	0.00209876543209877\\
0.624750881586509	0.000175604423868313\\
0.675629046610219	1.20536025667852e-05\\
0.72643579802999	1.75911852977688e-06\\
0.777175050783995	3.66637335679812e-07\\
};
\addlegendentry{Surrogate BER};

\addplot [color=green,dashed,line width=1.0pt,mark=triangle,mark options={solid}]
  table[row sep=crcr]{%
0.265931622599248	1\\
0.317525865515046	1\\
0.368997666751107	0.916666666666667\\
0.420353421431698	0.777777777777778\\
0.471599736825913	0.479166666666667\\
0.522743412656327	0.261904761904762\\
0.573791420494144	0.0833333333333333\\
0.624750881586509	0.0138888888888889\\
0.675629046610219	0.00188323917137476\\
0.72643579802999	0.000139988800895928\\
0.777175050783995	8.99928005759539e-05\\
};
\addlegendentry{Surrogate FER};

\end{axis}
\end{tikzpicture}%
 \caption{8-ASK, shaped inputs: Performance comparison of optimized rate 2/3 code and bit-mapping optimized DVB-S2 code. The horizontal axis denotes the gap to AWGN capacity as explained in~\eqref{eq:def_snr_gap}.}
 \label{fig:bmd8}
\end{figure}
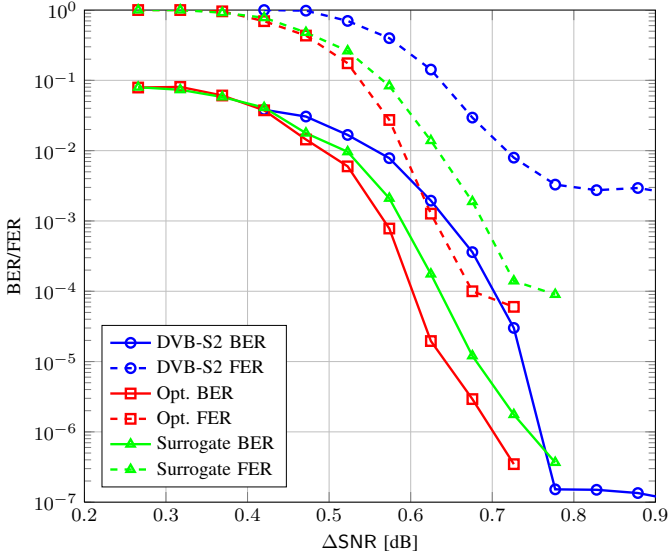

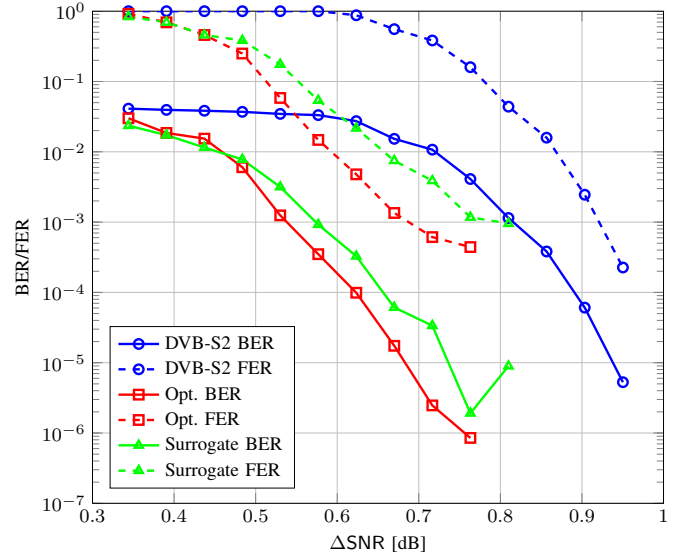
\begin{figure}
 \centering
 \footnotesize
%
%
\begin{tikzpicture}[scale=.9]

\begin{axis}[%
scale only axis,
xmin=0.3,
xmax=1,
xlabel={$\Delta\SNR$ [dB]},
xmajorgrids,
xminorgrids,
ymode=log,
ymin=1e-07,
ymax=1,
yminorticks=true,
ylabel={BER/FER},
ymajorgrids,
legend style={at={(0.03,0.03)},anchor=south west,draw=black,fill=white,legend cell align=left}
]
\addplot [color=blue,solid,line width=1.0pt,mark=o]
  table[row sep=crcr]{%
0.344095163934284	0.0410601851851852\\
0.390668588576453	0.0394814814814815\\
0.437198413770727	0.0383858024691358\\
0.48364726878372	0.0369976851851852\\
0.530147023895729	0.0345964506172839\\
0.576769188453206	0.0332577160493827\\
0.623352759992727	0.0273047839506173\\
0.669953531378109	0.0152824074074074\\
0.716581945229116	0.0107259259259259\\
0.763240633078119	0.00410069444444445\\
0.809928969148565	0.00114429012345679\\
0.856654074610837	0.000382010582010582\\
0.903420465565411	6.07389251997095e-05\\
0.950231243888531	5.2818818366082e-06\\
};
\addlegendentry{DVB-S2 BER};

\addplot [color=blue,dashed,line width=1.0pt,mark=o,mark options={solid}]
  table[row sep=crcr]{%
0.344095163934284	1\\
0.390668588576453	1\\
0.437198413770727	1\\
0.48364726878372	1\\
0.530147023895729	1\\
0.576769188453206	1\\
0.623352759992727	0.875\\
0.669953531378109	0.555555555555556\\
0.716581945229116	0.383333333333333\\
0.763240633078119	0.159722222222222\\
0.809928969148565	0.043859649122807\\
0.856654074610837	0.0158730158730159\\
0.903420465565411	0.00245098039215686\\
0.950231243888531	0.000226142017186793\\
};
\addlegendentry{DVB-S2 FER};

\addplot [color=red,solid,line width=1.0pt,mark=square]
  table[row sep=crcr]{%
0.344095163934284	0.0299884259259259\\
0.390668588576453	0.0185411522633745\\
0.437198413770727	0.0153927469135802\\
0.48364726878372	0.00603284832451499\\
0.530147023895729	0.00124598765432099\\
0.576769188453206	0.000349639462471321\\
0.623352760076759	9.88671336092539e-05\\
0.669953531378109	1.73948220064725e-05\\
0.716581945229112	2.46890919308932e-06\\
0.763240633247293	8.51750443253711e-07\\
};
\addlegendentry{Opt. BER};

\addplot [color=red,dashed,line width=1.0pt,mark=square,mark options={solid}]
  table[row sep=crcr]{%
0.344095163934284	0.916666666666667\\
0.390668588576453	0.694444444444444\\
0.437198413770727	0.458333333333333\\
0.48364726878372	0.25\\
0.530147023895729	0.0583333333333333\\
0.576769188453206	0.0147492625368732\\
0.623352760076759	0.00477554918815664\\
0.669953531378109	0.00134843581445523\\
0.716581945229112	0.000611845325501713\\
0.763240633247293	0.000441618088676912\\
};
\addlegendentry{Opt. FER};

\addplot [color=green,solid,line width=1.0pt,mark=triangle]
  table[row sep=crcr]{%
0.344095163934284	0.0235640432098765\\
0.390668588576453	0.0172458847736626\\
0.437198413770727	0.0115015432098765\\
0.48364726878372	0.00778086419753086\\
0.530147023895729	0.00316033950617284\\
0.576769188453206	0.00092189366786141\\
0.623352759992727	0.000327180535513869\\
0.669953531378109	6.10571333665504e-05\\
0.716581945229116	3.3652741889176e-05\\
0.763240633247293	1.91007393555313e-06\\
0.809928969148565	8.94390383514596e-06\\
};
\addlegendentry{Surrogate BER};

\addplot [color=green,dashed,line width=1.0pt,mark=triangle,mark options={solid}]
  table[row sep=crcr]{%
0.344095163934284	0.833333333333333\\
0.390668588576453	0.694444444444444\\
0.437198413770727	0.458333333333333\\
0.48364726878372	0.383333333333333\\
0.530147023895729	0.175\\
0.576769188453206	0.0537634408602151\\
0.623352759992727	0.0216450216450216\\
0.669953531378109	0.00747384155455904\\
0.716581945229116	0.00387596899224806\\
0.763240633247293	0.00116877045348294\\
0.809928969148565	0.000954016409082236\\
};
\addlegendentry{Surrogate FER};

\end{axis}
\end{tikzpicture}%
 \caption{64-ASK, shaped inputs: Performance comparison of rate 5/6 code and bit-mapping optimized DVB-S2 code. The horizontal axis denotes the gap to AWGN capacity as explained in~\eqref{eq:def_snr_gap}.}
 \label{fig:bmd64}
\end{figure}

As for the uniform case, the optimized protograph basematrices have been lifted yielding quasi-cyclic parity-check matrices of blocklengths $n=64800$.
100 decoding iterations have been performed. The simulation results are depicted in Fig.~\ref{fig:bmd8} and \ref{fig:bmd64}: At a target FER of $10^{-3}$,
the rate 2/3 code has a gap of \SI{0.63}{\decibel} to the AWGN capacity $\frac{1}{2}\log_2(1+\SNR)$ at a spectral efficiency of 1.38 bits/channel use. For the 5/6 rate code, we observe a gap of \SI{0.69}{\decibel} compared to AWGN capacity at a target
FER of $10^{-3}$ and a spectral efficiency of 4.25 bits/channel use. Hence, the bit-mapping optimized DVB-S2 codes are outperformed by \SI{0.1}{\decibel} (2/3 code) and \SI{0.25}{\decibel} (5/6 code).

\section{Conclusion}
\label{sec:conclusion}

We proposed a protograph-based LDPC code design approach for 
bandwidth-efficient coded modulation that is suitable both for uniform 
and shaped inputs. The different bit-channels are replaced 
by biAWGN surrogates so that PEXIT and differential evolution give ensembles with 
good decoding thresholds. The performance of the new codes for uniform inputs are as good as the best codes in literature. 
For shaped inputs, the new codes operate within \SI{0.7}{\decibel} to $\frac{1}{2}\log_2(1+\SNR)$. Future research should investigate the influence 
of the surrogates on the code performance by employing a full-fledged density evolution for protographs. Furthermore, precoded protographs 
should be considered to improve the threshold without increasing the variable node degrees.

\newcommand{\BIBdecl}{\setlength{\itemsep}{0.01em}}
\bibliographystyle{IEEEtran}
\normalsize
\bibliography{IEEEabrv,confs-jrnls,literature}

\begin{thebibliography}{10}
\providecommand{\url}[1]{#1}
\csname url@samestyle\endcsname
\providecommand{\newblock}{\relax}
\providecommand{\bibinfo}[2]{#2}
\providecommand{\BIBentrySTDinterwordspacing}{\spaceskip=0pt\relax}
\providecommand{\BIBentryALTinterwordstretchfactor}{4}
\providecommand{\BIBentryALTinterwordspacing}{\spaceskip=\fontdimen2\font plus
\BIBentryALTinterwordstretchfactor\fontdimen3\font minus
  \fontdimen4\font\relax}
\providecommand{\BIBforeignlanguage}[2]{{%
\expandafter\ifx\csname l@#1\endcsname\relax
\typeout{** WARNING: IEEEtran.bst: No hyphenation pattern has been}%
\typeout{** loaded for the language `#1'. Using the pattern for}%
\typeout{** the default language instead.}%
\else
\language=\csname l@#1\endcsname
\fi
#2}}
\providecommand{\BIBdecl}{\relax}
\BIBdecl

\bibitem{zehavi_bicm}
E.~Zehavi, ``8-{PSK} trellis codes for a {R}ayleigh channel,'' \emph{{IEEE}
  Trans. Commun.}, vol.~40, no.~5, pp. 873--884, May 1992.

\bibitem{caire_bicm}
G.~Caire, G.~Taricco, and E.~Biglieri, ``Bit-interleaved coded modulation,''
  \emph{{IEEE} Trans. Inf. Theory}, vol.~44, no.~3, pp. 927--946, 1998.

\bibitem{martinez_mismatched}
A.~Martinez, A.~Guillen~i Fabregas, G.~Caire, and F.~Willems, ``Bit-interleaved
  coded modulation revisited: A mismatched decoding perspective,'' \emph{{IEEE}
  Trans. Inf. Theory}, vol.~55, no.~6, pp. 2756--2765, June 2009.

\bibitem{fabregas_bicm_shaping}
A.~Guill\'en~i F\`abregas and A.~Martinez, ``Bit-interleaved coded modulation
  with shaping,'' in \emph{IEEE Inf. Theory Workshop (ITW)}, 2010.

\bibitem{boecherer_bmd}
G.~B\"ocherer, ``Probabilistic signal shaping for bit-metric decoding,'' in
  \emph{Proc. IEEE Int. Symp. Inf. Theory (ISIT)}, June 2014, pp. 431--435.

\bibitem{bocherer2014achievable}
\BIBentryALTinterwordspacing
------, ``Achievable rates for shaped bit-metric decoding,'' \emph{arXiv
  preprint}, 2015. [Online]. Available: \url{http://arxiv.org/abs/1410.8075}
\BIBentrySTDinterwordspacing

\bibitem{li2005}
Y.~Li and W.~Ryan, ``Bit-reliability mapping in {LDPC}-coded modulation
  systems,'' \emph{{IEEE} Commun. Lett.}, vol.~9, no.~1, pp. 1--3, Jan 2005.

\bibitem{lei2008}
J.~Lei and W.~Gao, ``Matching graph connectivity of {LDPC} codes to high-order
  modulation by bit interleaving,'' in \emph{Proc. Allerton Conf. Commun.,
  Contr., Comput.}, Sept 2008, pp. 1059--1064.

\bibitem{hager2014optimized}
C.~H\"ager, A.~Graell~i Amat, A.~Alvarado, F.~Br\"annstr\"om, and E.~Agrell,
  ``Optimized bit mappings for spatially coupled {LDPC} codes over parallel
  binary erasure channels,'' in \emph{Proc. IEEE Int. Conf. Commun. (ICC)},
  Jun. 2014, pp. 2064--2069.

\bibitem{zhang_MET_journal}
L.~Zhang and F.~Kschischang, ``Multi-edge-type low-density parity-check codes
  for bandwidth-efficient modulation,'' \emph{{IEEE} Trans. Commun.}, vol.~61,
  no.~1, pp. 43--52, January 2013.

\bibitem{richardson_MET}
T.~Richardson and R.~Urbanke, ``Multi-edge type {LDPC} codes,'' \emph{Workshop
  honoring Prof. Bob McEliece on his 60th birthday, California Institute of
  Technology, Pasadena, California}, pp. 24--25, 2002.

\bibitem{ten_brink_exit1}
S.~ten Brink, ``Convergence of iterative decoding,'' \emph{Electron. Lett.},
  vol.~35, no.~10, pp. 806--808, 1999.

\bibitem{thorpe_protograph}
J.~Thorpe, ``Low-density parity-check ({LDPC}) codes constructed from
  protographs,'' \emph{IPN progress report}, vol.~42, no. 154, pp. 42--154,
  2003.

\bibitem{divsalar_error_floor_cm}
D.~Divsalar and C.~Jones, ``Protograph based low error floor {LDPC} coded
  modulation,'' in \emph{IEEE Mil. Commun. Conf. (MILCOM)}, Oct 2005, pp.
  378--385 Vol. 1.

\bibitem{marinoni2010proto}
A.~Marinoni, P.~Savazzi, and R.~Wesel, ``Protograph-based q-ary {LDPC} codes
  for higher-order modulation,'' in \emph{Proc. Int. Symp. Turbo Codes and
  Iterative Inf. Process. (ISTC)}, Sep. 2010, pp. 68--72.

\bibitem{liva_pexit}
G.~Liva and M.~Chiani, ``Protograph {LDPC} code design based on {EXIT}
  analysis,'' in \emph{IEEE Global Telecommun. Conf. (GLOBECOM)}, 2007, pp.
  3250--3254.

\bibitem{gray1953pulse}
F.~Gray, ``Pulse code communication,'' Mar.~17 1953, {US} Patent 2,632,058.

\bibitem{mct}
T.~Richardson and R.~Urbanke, \emph{Modern Coding Theory}.\hskip 1em plus 0.5em
  minus 0.4em\relax Cambridge Univ. Press, 2008.

\bibitem{fabregas_ft}
A.~Guill\'en~i F\`abregas, A.~Martinez, and G.~Caire, ``Bit-interleaved coded
  modulation,'' \emph{Found. Trends Comm. Inf. Theory}, vol.~5, no. 1--2, pp.
  1--153, 2008.

\bibitem{schulte2015constant}
\BIBentryALTinterwordspacing
P.~Schulte and G.~B\"ocherer, ``Constant composition distribution matching,''
  \emph{arXiv preprint}, 2015. [Online]. Available:
  \url{http://arxiv.org/abs/1503.05133}
\BIBentrySTDinterwordspacing

\bibitem{bocherer2015bandwidth}
\BIBentryALTinterwordspacing
G.~B\"ocherer, P.~Schulte, and F.~Steiner, ``Bandwidth efficient and
  rate-matched low-density parity-check coded modulation,'' \emph{arXiv
  preprint}, 2015. [Online]. Available: \url{http://arxiv.org/abs/1502.02733}
\BIBentrySTDinterwordspacing

\bibitem{boecherer_thesis}
G.~B\"ocherer, ``Capacity-achieving probabilistic shaping for noisy and
  noiseless channels,'' Ph.D. dissertation, RWTH Aachen University, 2012.

\bibitem{Liva2013:Proto_TWCOM}
P.~Pulini, G.~Liva, and M.~Chiani, ``Unequal diversity {LDPC} codes for relay
  channels,'' \emph{IEEE Trans. Wireless Commun.}, vol.~12, no.~11, pp.
  5646--5655, Nov. 2013.

\bibitem{richardson_design_irregular}
T.~J. Richardson, M.~A. Shokrollahi, and R.~L. Urbanke, ``Design of
  capacity-approaching irregular low-density parity-check codes,'' \emph{{IEEE}
  Trans. Inf. Theory}, vol.~47, no.~2, pp. 619--637, 2001.

\bibitem{storn_DE}
R.~Storn and K.~Price, ``Differential evolution--a simple and efficient
  heuristic for global optimization over continuous spaces,'' \emph{Journal of
  global optimization}, vol.~11, no.~4, pp. 341--359, 1997.

\end{thebibliography}

\end{document}